\documentclass[12pt]{article}
\usepackage{epsfig,amsfonts,amssymb,}
\usepackage{hyperref}
\usepackage{mathtools,slashed}
\usepackage{cite}
\usepackage{braket}
\usepackage{amsmath, leftidx}
\usepackage{cite}
\usepackage{comment}
\usepackage{IEEEtrantools}
\usepackage{graphicx}

\usepackage[dvipsnames]{xcolor}
\usepackage{tikz}
\usetikzlibrary{shapes.misc}

\def\ZZZ{{\hbox{ Z\kern-1.6mm Z}}}
\def\RRR{{\hbox{ R\kern-2.4mm R}}}
\def\CCC{{\hbox{ C\kern-2.0mm C}}}
\def\zzz{{\hbox{z\kern-1mm z}}}

\newcommand{\qeq}{{\hbox{=\kern-2.3mm ? \kern.5mm }}}
\renewcommand{\qeq}{=}

\newcommand{\eps}{\epsilon}

\newcommand{\be}{\begin{equation}}
\newcommand{\ee}{\end{equation}}
\newcommand{\ben}{\begin{eqnarray}\displaystyle}
\newcommand{\een}{\end{eqnarray}}

\newcommand{\p}{\partial}

\def\one{{\hbox{ 1\kern-.8mm l}}}
\def\zero{{\hbox{ 0\kern-1.5mm 0}}}

\newcommand{\bea}[1]{\begin{eqnarray}\label{#1} }
\newcommand{\eea}{\end{eqnarray}}

\usepackage{bm}

\begin{document}
\begin{center}

{\Large \bf Classical spinning soft factors from gauge theory amplitudes}

\end{center}

\vskip .6cm
\medskip

\vspace*{4.0ex}

\baselineskip=18pt

\centerline{\large \rm Manu A$^{a,c}$, Debodirna Ghosh$^{b,c}$}

\vspace*{4.0ex}

\centerline{\large \it ~$^a$Institute of Physics, Sachivalaya Marg}
\centerline{\large \it  Bhubaneswar, Odisha - 751005, India}
\vspace{0.5cm}

\centerline{\large \it ~$^b$The Institute of Mathematical Sciences}
\centerline{\large \it IV Cross Road, C.I.T. Campus}
\centerline{\large \it Taramani, Chennai - 600113, India}
\vspace{0.5cm}

\centerline{\large \it and}
\vspace{0.5cm}

\centerline{\large \it ~$^c$Homi Bhabha National Institute}
\centerline{\large \it Training School Complex, Anushakti Nagar,}
\centerline{\large \it  Mumbai - 400085, India}

\vspace*{1.0ex}
\centerline{\small E-mail:  manu.akavoor@gmail.com, debodirna91@gmail.com}

\vspace*{5.0ex}

\centerline{\bf Abstract} \bigskip

In this short note, we analyse low energy electromagnetic radiation for spinning particles using the KMOC formalism\cite{MOV} and the quantum soft theorems. In particular, we study low energy electromagnetic radiation emitted by the so-called $\sqrt{\text{Kerr}}$ object. The $\sqrt{\text{Kerr}}$ is a solution of the free Maxwell's equations with infinite multipole moments expressed solely in terms of the charge(Q), mass(m) and spin(S) of the classical object . We consider the scattering of two $\sqrt{\text{Kerr}}$ particles and using the KMOC formalism generalised to spinning particles, we perturbatively prove the classical subleading soft photon theorem to leading order in the low deflection parameter and to $\mathcal{O}(S_{1}S_{2})$ in spin.



\vfill \eject

\baselineskip 18pt

\tableofcontents

\baselineskip 18pt

\section{Introduction}

Classical soft theorems are exact statements about the long wavelength radiation emitted in any classical scattering process. The theorems state that in spacetimes greater than four ($D>4$) the radiative gauge / gravity field measured at large distances is proportional to the ``classical limit" of the single soft factor with the momentum and angular momentum operators replaced with their classical asymptotic values. They have been derived from the quantum soft theorems using the saddle point approximation\cite{laddha18} as well as have been given a classical derivation using classical equations of motion, which holds for generic classical scattering processes\cite{laddha19}\footnote{The derivation from quantum soft theorems is valid for certain classes of scattering processes which can be characterised as large impact parameter scattering (low deflection processes) and the probe-scatterer approximation (the ratio of the probe to scatterer masses is small).}.\\
In $D=4$ the situation is more complicated but more intriguing due to the presence of infrared divergences in the $S$-matrix. The (quantum) soft theorems receive non-analytic corrections starting from the subleading term, $\ln\omega$\cite{sahoo18,sahoo20}. Classical low energy gravitational (and electromagnetic) radiation have been analysed for generic classical scattering processes using the classical equations of motion and consequently, classical soft theorems in $D=4$ have been derived\cite{sahoo19,sahoo21}. They hold for spinning particles as well as scalars. Unlike the situation in $D>4$, the proof of the classical soft theorems in $D=4$ from their quantum counterparts is lacking. Given the developments in the literature on computing classical observables from scattering amplitudes, one can hope to prove the classical soft theorems in a perturbative setting. 
For classical large impact parameter scattering, the first steps towards this goal was undertaken in \cite{athira20}. For such scattering processes, using the KMOC formalism\cite{KMOC} and (quantum) subleading soft theorems in $D=4$, the low energy gravitational and electromagnetic radiation was analysed and subsequently, the classical soft theorems for scalar particles were proved till Next-to-Leading order (NLO) in the coupling (low deflection parameter)\footnote{Recently, the classical leading soft factor in electrodynamics has been proved till NLO\cite{laddha21}.}.\\
The KMOC formalism is a framework which computes classical observables like linear impulse, directly using on-shell scattering amplitudes. The basic idea is to start with an initial coherent state of particles, with momenta peaked around their classical values, in the far past, evolve it with the $S$-matrix and then take the expectation value of a quantum observable. Classical limit of the observable then gives the classical result. The small coupling expansion of the scattering amplitude, in the classical limit, is then identified as the large impact parameter expansion. The KMOC formalism has now been generalised to include asymptotic states like spinning particles, particles with colour as well as incoming radiation \cite{MOV,delacruz20,ochirov21,kosower21}. The formalism is one of the multiple methods used to compute classical observables from scattering amplitudes, relevant for the binary black hole system. For a selection of such results we refer the reader to\cite{portok,portok1,portok2,cfq1,cfq2,cfq2.4,cfq2.5,cfq3,cfq3.5,cfq4,cfq4.5,cfq5,cfq6,cfq7,cfq8,bern1,bern2,bern3,bern4}.\\
In this short note, we take the first steps to provide a perturbative proof of the classical subleading soft theorems for spinning particles, from the quantum subleading soft theorems. For this paper, we shall concentrate on electromagnetic radiation and to Leading order (LO) in the low deflection parameter. We will concentrate on the electromagnetic analog of the Kerr black hole, the so-called $\sqrt{\text{Kerr}}$ particle\footnote{We shall provide the relevant details of this classical object later in the paper.}. Since, spinning particles have internal angular momentum there is another length scale which is called the spin expansion parameter, denoted by the spin of the particle(S). Our main result is thus the computation of the radiative gauge field for soft radiation to subleading order in the soft expansion for $\sqrt{\text{Kerr}_{1}}-\sqrt{\text{Kerr}_{2}}$ scattering upto $\mathcal{O}(S_{1}S_{2})$ and leading order in the low deflection parameter.\\
This paper is organised as follows. In sect.\ref{sec2} we use the classical equations of motion to derive the the soft radiative gauge field for $\sqrt{\text{Kerr}_{1}}-\sqrt{\text{Kerr}_{2}}$ scattering till leading order in the coupling and $\mathcal{O}(S_{1}S_{2})$. In sect.\ref{sec3} we give a brief summary of the KMOC and its generalisation to spinning particles and in sect.\ref{sec5} we compute the radiative gauge field created by two scattering $\sqrt{\text{Kerr}}$s using the KMOC formalism to leading order int the coupling and $\mathcal{O}(S_{1}S_{2})$, thus verifying the classical results obtained. We end the paper with a few remarks and some open questions.

\section{Radiative gauge field from classical dynamics in the soft limit}\label{sec2}

In this section we shall compute the radiative gauge field for low frequency radiation in the scattering process of $\sqrt{\text{Kerr}_{1}}-\sqrt{\text{Kerr}_{2}}$. Before that we shall review the necessary features of classical spinning particles relevant for us and also give a brief summary of the $\sqrt{\text{Kerr}}$ solution.\\
Classical spinning particles in addition to mass also have internal spin degrees of freedom which in the non - relativistic case can be described by the internal coordinates fixed to the body, for eg. Euler angles. In the covariant description
the internal spin of a classical spinning particle can either be described in terms of an anti-symmetric $2$-tensor, $S^{\mu\nu}$ or a $4$-vector, $s^{\mu}$. Since soft theorems naturally involve the angular momentum tensor, we shall be utilising the anti-symmetric $2$- tensor, $S_{\mu\nu}$. However, we note that $S^{\mu\nu}$ has 6 components whereas we know that in the rest frame of the particle, we can describe the spin in terms of the rotations about the three spatial axes. Hence, to obtain the correct physical degrees of freedom, we impose the spin supplementary condition (SSC)
\begin{equation}
S^{\mu\nu}(\tau)p_{1\nu}(\tau) = 0.
\end{equation}
Here $\tau$ is the proper time along the path of the particle and the SSC has to hold throughout the path of the particle. Also, the spin vector $s^{\mu}$ and $S^{\mu\nu}$ are related by
\begin{equation}\label{eq:22}
S^{\mu\nu} =\frac{1}{m}\epsilon^{\mu\nu\rho\sigma}p_{\rho}s_{\sigma},\ \ \ \ \ \ \ \ \ s^{\mu} = \frac{1}{2m}\epsilon^{\mu\nu\rho\sigma}p_{\nu}S_{\rho\sigma}.
\end{equation}
We shall be analysing low frequency radiation for large impact parameter scattering. To derive the equations of motion for the $\sqrt{\text{Kerr}}$ till leading order in the low deflection parameter, we need its conserved current. This can be obtained by doing the single copy substitution on the conserved stress tensor for the Kerr black hole\footnote{There is a hierarchy of substitutions which can be used to go between classical solutions of Bi-adjoint scalar theory, gauge theories and gravitational theories\cite{bernckreview}. The inverse map from gravitational solutions to gauge theory solutions is generally known as single copy.}. For the gravity case, the low deflection parameter in $D=4$ is given by
\begin{equation}
\frac{G_{N}E}{|\vec{b}|}\ll 1.
\end{equation}
From the Effective field theory perspective, the dynamical equations for spinning particles interacting via gravity have been derived in \cite{porto05,levi15}. For the special case of Kerr black holes, to leading order in the low deflection parameter the equations can be obtained by using the conserved stress tensor\cite{vines}
\begin{equation}\label{stgr}
T^{\mu\nu}(x) = \frac{1}{m}\int d\tau p^{(\mu} \ p^{\rho}(\tau)\ \exp{(a\star \p)}^{\nu)}_{\ \rho}\ \hat{\delta}^{(4)}(x-r(\tau)).
\end{equation}
Here $a^{\mu}$ is the internal spin of the black hole and it is related to the pseudo - spin vector as $a^{\mu} = s^{\mu}/m$. Also $(a\star \p)^{\mu}_{\ \rho} =\epsilon^{\mu}_{\ \rho\alpha\beta}a^{\alpha}\frac{\p}{\p x_{\beta}}$. From the above we can obtain the conserved current for the $\sqrt{\text{Kerr}}$ by making the single copy substitution, $p_{\mu}\rightarrow Q$. We get
\begin{equation}\label{eq:24}
J^{\mu}(x) = \frac{Q}{m}\int d\tau\ p^{\rho}(\tau)\ \exp{(a\star \p)}^{\mu}_{\ \rho}\ \hat{\delta}^{(4)}(x-r(\tau))
\end{equation}
This is the conserved current for the $\sqrt{\text{Kerr}}$ object to leading order in the low deflection parameter and it is characterised by the charge $Q$, mass $m$ and the spin, $a^{\mu} = s^{\mu}/m$. To cross check if this is the correct conserved current for the $\sqrt{\text{Kerr}}$ particle, we will evaluate the electrostatic and magnetostatic potential obtained from this current to $\mathcal{O}(a^{2})$ and match it to the one obtained by doing the complex deformation of the Coulomb solution\cite{lynden-bell}. The gauge field can be calculated from the general solution of the Maxwell solution
\begin{equation}
A^{\mu}(x) = \int d^{4}x'\ G_{r}(x-x')\ J^{\mu}(x')
\end{equation}
where $G_{r}(x-x') = \delta \big(\frac{(x-x')^{2}}{2}\big)\Theta(x-x')$ is the standard retarted Green's function. Plugging eq.\eqref{eq:24} in the above equation, expanding the exponential and doing by-parts we obtain the gauge field,
\begin{equation}
A^{\mu}(x) = \frac{Q}{m}\int d^{4}x'\ \delta^{(4)}(x'-r(\tau))\bigg[ \delta^{\mu}_{\rho} + (a\star \p)^{\mu}_{\ \rho} + \frac{1}{2!}(a\star \p)^{\mu}_{\ \alpha}(a\star \p)^{\alpha}_{\ \rho}\ +\ldots\bigg]\frac{p^{\rho}}{r}
\end{equation}
The above calculation is done along the same lines as the Kerr black hole in \cite{vines}. We can now do the remaining integrals by going to the rest frame of the particle, $p^{\mu} = (m,\vec{0})$ and putting $r^{\mu}(\tau) = \tau$. The SSC in terms of the spin psuedo-vector is $a\cdot p = 0$ which in the rest frame gives $a^{0} = 0$. The zeroth component of the gauge field is then
\begin{equation}
A^{0}(x) = \phi(x) = Q\bigg( 1 - \frac{(\vec{a}\cdot \vec{\p})^{2}}{2!} +\ldots\bigg)\frac{1}{r}
\end{equation} 
and the vector potential is 
\begin{equation}
A^{i}(x) = Q\eps^{ijk}a^{j}\p_{k}\bigg(1 -\frac{(\vec{a}\cdot \vec{\p})^{2}}{3!} +\ldots\bigg)\frac{1}{r}
\end{equation}
The derivatives can be evaluated straightforwardly and we get
\begin{equation}
\phi(x) = \frac{Q}{r}\bigg( 1 - \frac{1}{2!}\bigg( -\frac{\vec{a}^{2}}{r^{2}} + 3\frac{(\vec{a}\cdot \vec{x})^{2}}{r^{4}}\bigg)+\ldots\bigg),\ \ \ \ \ 
A^{i}(x) = Q\ \eps^{ijk}a^{j}\bigg(-\frac{x_{k}}{r^{3}}+\ldots\bigg)
\end{equation}
To obtain the magnetostatic potential\footnote{In regions where there are no sources, the curl of the magnetic field is also zero which allows us to define a magnetostatic potential, $B_{i}(x) = \p_{i}\chi$.} we calculate the magnetic field $B_{i} = (\nabla \times \vec{A})_{i}$ and from it derive the magnetostatic potential defined as $B_{i} = \p_{i}\chi$. Doing this computation, we get the magnetic field to be
\begin{equation}
B_{i}(x) = -Q\frac{a_{i}}{r^{3}} + 3Q\ x_{i}\frac{\vec{a}\cdot \vec{x}}{r^{5}}+\ldots = \p_{i}\bigg( -Q\ \frac{\vec{a}\cdot \vec{x}}{r^{3}}+\ldots\bigg)
\end{equation}
Hence, from the conserved current we get the potentials to be 
\begin{equation}\label{eq:212}
\phi(x) = \frac{Q}{r}\bigg( 1 - \frac{1}{2!}\bigg( -\frac{\vec{a}^{2}}{r^{2}} + 3\frac{(\vec{a}\cdot \vec{x})^{2}}{r^{4}}\bigg)+\ldots\bigg),\ \ \ \ \ \chi = -Q\ \frac{\vec{a}\cdot \vec{x}}{r^{3}} +\ldots
\end{equation}
Now as explained in \cite{lynden-bell} the potential created by the $\sqrt{\text{Kerr}}$ object can also be obtained by doing a complex deformation of the Coulomb solution
\begin{equation}
\Phi(x) = \frac{Q}{r} \rightarrow \frac{Q}{\sqrt{(\vec{x}-i\vec{a})^{2}}} = \phi +i\chi.
\end{equation}
Expanding the denominator in the $r = \sqrt{\vec{x}^{2}}\gg |\vec{a}|$ regime, we get the above potential to be
\begin{equation}
\Phi(x) = \frac{Q}{r}\bigg( 1 +\frac{1}{2}\frac{\vec{a}^{2}}{r^{2}} - i\frac{\vec{a}\cdot \vec{x}}{r^{2}} - \frac{3}{2}\frac{(\vec{a}\cdot \vec{x})^{2}}{r^{4}} +\ldots\bigg)
\end{equation} 
From the above equation we can read off the electromagnetic and magnetostatic potential and they are
\begin{equation}
\phi(x) = \frac{Q}{r} +\ \frac{Q}{2}\frac{\vec{a}^{2}}{r^{3}}\ - \frac{3}{2}\frac{(\vec{a}\cdot \vec{x})^{2}}{r^{5}} +\ldots,\ \ \ \chi(x) = - Q\frac{\vec{a}\cdot \vec{x}}{r^{3}}+\ldots
\end{equation}
Hence, we see that there is an exact match with the same quantity calculated using the conserved current in eq.\eqref{eq:24}. For more details on the $\sqrt{\text{Kerr}}$ object we refer the reader to \cite{lynden-bell}.\\
We will be interested in the spin expansion till $\mathcal{O}(S)$, so, we expand the conserved current in eq.\eqref{eq:24} till $\mathcal{O}(a)$ to get
\begin{equation}\label{eq:25}
\begin{split}
J^{\mu}(x) &= \ \frac{Q}{m}\int d\tau\  p^{\mu}(\tau)\ \hat{\delta}^{(4)}(x-r(\tau))\ +\frac{Q}{m}\int d\tau\ (a\star \p)^{\mu}_{\ \rho}\ p^{\rho}\ \hat{\delta}^{(4)}(x-r(\tau)) +\mathcal{O}(a^{2})\\
& = Q\int d\tau\ v^{\mu}(\tau)\ \hat{\delta}^{(4)}(x-r(\tau))\ + \frac{Q}{m}\int d\tau\ S^{\mu\rho}\p_{\rho}\ \hat{\delta}^{(4)}(x-r(\tau)) +\mathcal{O}(S^{2})
\end{split}
\end{equation}
In going from the first line to the second line, we have used the first equation in eq.\eqref{eq:22}. So, this is the current of $\sqrt{\text{Kerr}}$ till $\mathcal{O}(S)$. Due to the anti-symmetry of $S_{\mu\nu}$ the current is conserved. We can also compute the gyromagnetic ratio, $g$, of the spinning particle from the above current. We do this by taking the non-relativistic limit of the interaction term which we obtain
\begin{equation}
V = -\frac{Q}{m}\vec{S}\cdot \vec{B}
\end{equation}
From the above equation we see that the $\sqrt{\text{Kerr}}$ has $g=2$.\\
We can now compute the equations of motion by coupling the current to the electromagnetic field. The equation for the linear momentum of the particle comes out to be
\begin{equation}
\frac{dp^{\mu}(\tau)}{d\tau} = QF^{\mu\nu}(x(\tau))v_{\nu} + \frac{Q}{2m}\ S^{\rho\sigma}\p^{\mu}F_{\rho\sigma}(x(\tau))
\end{equation}
The equation for the spin can be derived by demanding that the SSC holds throughout the path of the particle, taking the derivative of it 
\begin{equation}
\begin{split}
& \dot{S}^{\mu\nu}p_{\nu} + S^{\mu\nu}\dot{p}_{\nu} = 0, \\
& \dot{S}^{\mu\nu}p_{\nu} + \frac{Q}{m}\ S^{\mu\nu}F_{\nu\alpha}p^{\alpha} = 0,\\
& \dot{S}^{\mu\nu}p_{\nu} + \frac{Q}{m}\ S^{\mu\alpha}F_{\alpha}^{\ \nu}p_{\nu} = 0,\\ 
& \bigg(\dot{S}^{\mu\nu} + \frac{2Q}{m}\ S^{[\mu\alpha}F_{\alpha}^{\ \nu]}\bigg)p^{\nu} = 0
\end{split}
\end{equation}
In the last step, we have anti-symmetrized in the $(\mu\nu)$ indices. Therefore, the equation of motion for the $\sqrt{\text{Kerr}}$ till $O(S)$ and LO is
\begin{align}
&\frac{dp^{\mu}(\tau)}{d\tau} = QF^{\mu\nu}(x(\tau))v_{\nu} + \frac{Q}{2m}\ S^{\rho\sigma}\p^{\mu}F_{\rho\sigma}(x(\tau)) \label{eq:210}\\
&\frac{dS^{\mu\nu}(\tau)}{d\tau} =\ -\frac{2Q}{m}\ F^{[\mu}_{\ \ \alpha}S^{\nu]\alpha}(\tau)\label{eq:214}
\end{align}
Using these equations, the radiative gauge field can be computed using the Fourier transform of the current,
\begin{equation}
A_{\mu}(x) = \frac{Q}{2\pi r}\int d\omega\ e^{-i\omega t}\ \tilde{J}_{\mu}(k)
\end{equation}
where $k=\omega(1,\hat{n}) = \omega(1,\frac{\vec{x}}{r})$.
The Fourier transform of the current is what we shall be computing using the equations of motion and also comparing to the ones obtained from the KMOC formalism. Till $\mathcal{O}(S)$ this is 
\begin{multline}\label{eq:220}
\tilde{J}^{\mu}(k) = iQ\ \int d\sigma\ e^{ik.x(\sigma)}\ \frac{1}{k\cdot v(\sigma)}\bigg[ a^{\mu}(\sigma) - \frac{k\cdot a(\sigma)}{k\cdot v(\sigma)}\ v^{\mu}(\sigma)\bigg]\\
+\frac{Q}{m}\int d\sigma\ e^{ik\cdot x(\sigma)}\frac{1}{k\cdot v(\sigma)}\bigg[\dot{S}^{\mu\nu}(\sigma) - \frac{k\cdot a(\sigma)}{k\cdot v(\sigma)}S^{\mu\nu}(\sigma)\bigg]k_{\nu} 
\end{multline}
Here $a^{\mu}(\sigma) =\frac{d^{2}z^{\mu}(\sigma)}{d\sigma^{2}}$ is the acceleration of the spinning particle, not be confused with the rescaled spin vector.\\
We shall be studying $\sqrt{\text{Kerr}_{1}}-\sqrt{\text{Kerr}_{2}}$ scattering in the low deflection regime (large impact parameter). This is the same class of scattering processes studied in \cite{gb1,gb2}. The setup is that initially in the far past the particles are assumed to be free, then they undergo large impact scattering and in the process they emit soft radiation. We take the path to be parametrised as
\begin{equation}
x^{\mu}_{1}(\tau) = b_{1}^{\mu} +v_{1}\tau +z_{1}^{\mu}(\tau),\ \ \ \ \ x_{2}^{\mu}(\tau) = v_{2}^{\mu}\tau + z_{2}^{\mu}(\tau)
\end{equation}
Here $b_{1}^{\mu}$ is the impact parameter, $v_{i}$s are the initial velocities and $z_{i}$s denote the deflection from the initial path. The spin angular momentum will be parametrised as
\begin{equation}
S^{\mu\nu}_{i}(\tau) = S^{\mu\nu}_{i} + \tilde{S}^{\mu\nu}_{i}(\tau)
\end{equation}
where $S_{i}^{\mu\nu}(\tau)$ is the initial spin of the two $\sqrt{\text{Kerr}}$ particles and $\tilde{S}_{i}^{\mu\nu}$ denotes the change in the initial spin angular momentum of the particles. Since the spinning object has an intrinsic length scale, there will be two dimensionless parameters. These are
\begin{equation}
\frac{e^{2}}{Eb} \ll 1\ \ ; \ \ \frac{e^{2}a_{i}}{Eb^{2}} \ll 1.
\end{equation}
 Here $b$ is the impact parameter of the scattering process, $E$ is the energy of the massive spinning particle, $a_{i}$s are the characteristic length for the $\sqrt{\text{Kerr}}$ objects and $e$ is the coupling constant. The first inequality is obtained by requiring that the deflection should be much less than the impact parameter and the second comes from the requirement that the change in '$a$' be much smaller than the impact parameter. Additionally, it can be checked that requiring low deflection ensures that the change in '$a$' is much smaller than '$a$'.\\

\subsection{Soft Radiative gauge field for $\sqrt{\text{Kerr}_{1}}-\sqrt{\text{Kerr}_{2}}$}

In this subsection we compute the radiative gauge field in the soft limit for $\sqrt{\text{Kerr}_{1}}-\sqrt{\text{Kerr}_{2}}$ scattering process. We shall restrict ourselves to $\mathcal{O}(S_{1}S_{2})$ in the spin expansion and to leading order in the low deflection parameter.\\
Since the second particle is also a $\sqrt{\text{Kerr}}$, the EM field created by it also has a contribution coming from its spin angular momentum. So, the EM field 
\begin{equation}
F_{2}^{\mu\nu}(x) = iQ_{2}\int \frac{d^{4}l}{(2\pi)^{4}}\ \hat{\delta}(l\cdot v_{2})\ e^{-il\cdot x}\ \frac{l^{\mu}}{l^{2}}\ \big((l\wedge v_{2})^{\mu\nu} -\frac{i}{m_{2}}(l\wedge S_{2}^{-\alpha})^{\mu\nu}l_{\alpha}\big)
\end{equation}
Using this the acceleration of the first particle comes out to be
\begin{equation}\label{eq:228}
\begin{split}
a_{1}^{\mu}(\tau) & = \frac{iQ_{1}Q_{2}}{m_{1}}\int \frac{d^{4}l}{(2\pi)^{4}}\ \hat{\delta}(l\cdot v_{2})\ e^{-il\cdot b_{1}}\ e^{-i(l\cdot v_{1})\tau}\ \frac{1}{l^{2}}\\
&\bigg[\ l^{\mu}\bigg(v_{1}\cdot v_{2} -\frac{i}{m_{2}}(v_{1}\wedge l)_{2} -\frac{i}{m_{1}}(l\wedge v_{2})_{1} +\frac{(l\cdot S_{1}^{-\sigma})(l\cdot S^{-}_{{2}\sigma})}{m_{1}m_{2}}\bigg) - (l\cdot v_{1})(v_{2}^{\mu} -\frac{i}{m_{2}}S_{2}^{\mu\alpha}l_{\alpha})\bigg]
\end{split}
\end{equation}
Therefore the change in the linear impulse is given by
\begin{equation}\label{eq:229}
\Delta p_{1}^{\mu}(\tau) = iQ_{1}Q_{2}\int d\mu(1,2,l)\ e^{-il\cdot b_{1}}\ \frac{l^{\mu}}{l^{2}} \bigg(v_{1}\cdot v_{2} -\frac{i}{m_{2}}(v_{1}\wedge l)_{2} -\frac{i}{m_{1}}(l\wedge v_{2})_{1} +\frac{(l\cdot S_{1}^{-\sigma})(l\cdot S^{-}_{{2}\sigma})}{m_{1}m_{2}}\bigg)
\end{equation}
Similarly the computation of the change in the spin angular momentum of the first particle can also be computed and we obtain
\begin{equation}\label{eq:230}
\begin{split}
\dot{S}_{1}^{\mu\nu}(\tau) & = \frac{iQ_{1}Q_{2}}{m_{1}}\int \frac{d^{4}l}{(2\pi)^{4}}\ \hat{\delta}(l\cdot v_{2})\ e^{-il\cdot b_{1}}\ e^{-i(l\cdot v_{1})\tau}\ \frac{1}{l^{2}}\\
&\bigg[(v_{2}\wedge S_{1}^{-\alpha})^{\mu\nu}l_{\alpha} - (l\wedge S_{1}^{-\alpha})^{\mu\nu}v_{2\alpha} -\frac{i}{m_{2}}(l\wedge S_{1}^{-\alpha})^{\mu\nu}(l\cdot S^{-}_{2\alpha}) -\frac{i}{m_{2}}(S_{2}^{-\rho}\wedge S_{1}^{-\alpha})^{\mu\nu}\ l_{\rho}l_{\alpha}\bigg]
\end{split}
\end{equation}
The spin angular momentum impulse then becomes
\begin{equation}\label{eq:231}
\begin{split}
\Delta S_{1}^{\mu\nu}(\tau) & = \frac{iQ_{1}Q_{2}}{m_{1}}\int d\mu(1,2,l)\ e^{-il\cdot b_{1}}\ \frac{1}{l^{2}}\\
&\bigg[(v_{2}\wedge S_{1}^{-\alpha})^{\mu\nu}l_{\alpha} - (l\wedge S_{1}^{-\alpha})^{\mu\nu}v_{2\alpha} -\frac{i}{m_{2}}(l\wedge S_{1}^{-\alpha})^{\mu\nu}(l\cdot S^{-}_{2\alpha}) -\frac{i}{m_{2}}(S_{2}^{-\rho}\wedge S_{1}^{-\alpha})^{\mu\nu}\ l_{\rho}l_{\alpha}\bigg]
\end{split}
\end{equation}
Finally using eq.\eqref{eq:228} and eq.\eqref{eq:230}, we calculate the radiative gauge field in the soft limit in the same manner as in the previous sections. We get the radiative gauge field for the first $\sqrt{\text{Kerr}_{1}}$ for long wavelength at the subleading order in soft frequency to be
\begin{equation}
\begin{split}
\tilde{J}_{1}^{\mu}(k) &= \frac{iQ_{1}^{2}Q_{2}}{m_{1}}\int d\mu(1,2,l)\ e^{-il\cdot b_{1}}\ \frac{1}{l^{2}}\ \frac{b_{1}\cdot k}{k \cdot v_{1}}\ \frac{k_{\nu}}{k\cdot v_{1}}(v_{1}\wedge l)^{\mu\nu}\\
&\hspace{2cm}\bigg(v_{1}\cdot v_{2} -\frac{i}{m_{2}}(v_{1}\wedge l)_{2} -\frac{i}{m_{1}}(l\wedge v_{2})_{1} +\frac{(l\cdot S_{1}^{-\sigma})(l\cdot S^{-}_{{2}\sigma})}{m_{1}m_{2}}\bigg)\\
&\hspace{-1cm}-\frac{Q_{1}^{2}Q_{2}}{m_{1}}\int \frac{d^{4}l}{(2\pi)^{4}}\ \hat{\delta}'(l\cdot v_{1})\hat{\delta}(l\cdot v_{2})\frac{1}{l^{2}}\ e^{-il\cdot b_{1}}\ \frac{k_{\nu}}{k\cdot v_{1}}\ (v_{1}\wedge l)^{\mu\nu}\ \bigg((v_{1}\wedge l)_{2} -\frac{i}{m_{1}}(l\wedge v_{2})_{1} +\frac{(l\cdot S_{1}^{-\sigma})(l\cdot S^{-}_{{2}\sigma})}{m_{1}m_{2}}\bigg)\\
&+\frac{Q_{1}^{2}Q_{2}}{m_{1}}\int d\mu(1,2,l)\ e^{-il\cdot b_{1}}\ \frac{1}{l^{2}}\ \bigg[\bigg(v_{2}^{\mu} -\frac{k\cdot v_{2}}{k\cdot v_{1}}v_{1}^{\mu}\bigg) -\frac{i}{m_{2}}\bigg(S_{2}^{\mu\alpha}l_{\alpha} -\frac{(k\wedge l)_{2}}{k\cdot v_{1}} v_{1}^{\mu}\bigg)\bigg]\\
&+\frac{iQ_{1}^{2}Q_{2}}{m_{1}^{2}}\int d\mu(1,2,l) e^{-il\cdot b_{1}}\ \frac{1}{l^{2}}\\
& \frac{k_{\nu}}{k\cdot v_{1}}\bigg[(v_{2}\wedge S_{1}^{-\alpha})^{\mu\nu}l_{\alpha} - (l\wedge S_{1}^{-\alpha})^{\mu\nu}v_{2\alpha} -\frac{i}{m_{2}}(l\wedge S_{1}^{-\alpha})^{\mu\nu}(l\cdot S^{-}_{2\alpha}) -\frac{i}{m_{2}}(S_{2}^{-\rho}\wedge S_{1}^{-\alpha})^{\mu\nu}\ l_{\rho}l_{\alpha}\\
&\hspace{2cm}-\frac{k\cdot l}{k\cdot v_{1}}\bigg(v_{1}\cdot v_{2} -\frac{i}{m_{2}}(v_{1}\wedge l)_{2} -\frac{i}{m_{1}}(l\wedge v_{2})_{1} +\frac{(l\cdot S_{1}^{-\sigma})(l\cdot S^{-}_{{2}\sigma})}{m_{1}m_{2}}\bigg)S_{1}^{\mu\nu}\bigg]
\end{split}
\end{equation}
Using the change in linear impulse in eq.\eqref{eq:229} and change in spin angular momentum in eq.\eqref{eq:231}, the above expression can be written as
\begin{equation}\label{eq:438}
\begin{split}
\tilde{J}_{1}^{\mu}(k) &= \frac{Q_{1}}{m_{1}}\ \bigg(\frac{b_{1}\cdot k}{k\cdot v_{1}}\bigg)\ \frac{k_{\nu}}{k\cdot v_{1}}(v_{1}\wedge \Delta p_{1})^{\mu\nu} +\frac{Q_{1}}{m_{1}}\frac{k_{\nu}}{k\cdot v_{1}}\Delta S_{1}^{\mu\nu} - \frac{Q_{1}}{m_{1}^{2}}\ \frac{k_{\nu}}{(k\cdot v_{1})^{2}}(k\cdot \Delta p_{1})S_{1}^{\mu\nu}\\
& \hspace{-1cm}-\frac{Q_{1}^{2}Q_{2}}{m_{1}}\frac{k_{\nu}}{k\cdot v_{1}}\ \bigg(v_{1}\wedge \frac{\p}{\p v_{1}}\bigg)^{\mu\nu}\ \int d\mu(1,2,l)\ e^{-il\cdot b_{1}}\bigg(v_{1}\cdot v_{2} -\frac{i}{m_{2}}(v_{1}\wedge l)_{2} -\frac{i}{m_{1}}(l\wedge v_{2})_{1} +\frac{(l\cdot S_{1}^{-\sigma})(l\cdot S^{-}_{{2}\sigma})}{m_{1}m_{2}}\bigg)\\
\end{split}
\end{equation}
We shall compute the radiative gauge field emitted by the second $\sqrt{\text{Kerr}_{2}}$ particle. The EM field created by the first particle is
\begin{equation}
F_{1}^{\mu\nu}(x) = iQ_{1}\int \frac{d^{4}l}{(2\pi)^{4}}\ \hat{\delta}(l\cdot v_{1})\ e^{-il\cdot x}\ e^{il\cdot b_{1}}\ \frac{l^{\mu}}{l^{2}}\ \big((l\wedge v_{1})^{\mu\nu} -\frac{i}{m_{1}}(l\wedge S_{1}^{-\alpha})^{\mu\nu}l_{\alpha}\big)
\end{equation}
Using this the acceleration of the second particle comes out to be
\begin{equation}\label{eq:235}
\begin{split}
a_{2}^{\mu}(\tau) & = \frac{iQ_{2}Q_{1}}{m_{2}}\int \frac{d^{4}l}{(2\pi)^{4}}\ \hat{\delta}(l\cdot v_{1})\ e^{il\cdot b_{1}}\ e^{-i(l\cdot v_{2})\tau}\ \frac{1}{l^{2}}\\
&\bigg[\ l^{\mu}\bigg(v_{1}\cdot v_{2} -\frac{i}{m_{1}}(v_{2}\wedge l)_{1} -\frac{i}{m_{2}}(l\wedge v_{1})_{2} +\frac{(l\cdot S_{1}^{-\sigma})(l\cdot S^{-}_{{2}\sigma})}{m_{1}m_{2}}\bigg) - (l\cdot v_{2})(v_{1}^{\mu} -\frac{i}{m_{1}}S_{1}^{\mu\alpha}l_{\alpha})\bigg]
\end{split}
\end{equation}
Therefore the change in the linear impulse is given by
\begin{equation}\label{eq:236}
\Delta p_{2}^{\mu}(\tau) = -iQ_{1}Q_{2}\int d\mu(1,2,l)\ e^{-il\cdot b_{1}}\ \frac{l^{\mu}}{l^{2}} \bigg(v_{1}\cdot v_{2} -\frac{i}{m_{2}}(v_{1}\wedge l)_{2} -\frac{i}{m_{1}}(l\wedge v_{2})_{1} +\frac{(l\cdot S_{1}^{-\sigma})(l\cdot S^{-}_{{2}\sigma})}{m_{1}m_{2}}\bigg)
\end{equation}
Similarly the computation of the change in the spin angular momentum of the first particle can also be computed and we obtain
\begin{equation}\label{eq:237}
\begin{split}
\dot{S}_{2}^{\mu\nu}(\tau) & = -\frac{iQ_{1}Q_{2}}{m_{2}}\int \frac{d^{4}l}{(2\pi)^{4}}\ \hat{\delta}(l\cdot v_{1})\ e^{il\cdot b_{1}}\ e^{-i(l\cdot v_{2})\tau}\ \frac{1}{l^{2}}\\
&\bigg[(l\wedge S_{2}^{-\alpha})^{\mu\nu}v_{1\alpha} - (v_{1}\wedge S_{2}^{-\alpha})^{\mu\nu}l_{\alpha} +\frac{i}{m_{2}}(l\wedge S_{2}^{-\alpha})^{\mu\nu}(l\cdot S^{-}_{1\alpha}) +\frac{i}{m_{2}}(S_{1}^{-\rho}\wedge S_{2}^{-\alpha})^{\mu\nu}\ l_{\rho}l_{\alpha}\bigg]
\end{split}
\end{equation}
The spin angular momentum impulse then becomes
\begin{equation}\label{eq:238}
\begin{split}
\Delta S_{2}^{\mu\nu} & = \frac{iQ_{1}Q_{2}}{m_{2}}\int d\mu(1,2,l)\ e^{-il\cdot b_{1}}\ \frac{1}{l^{2}}\\
&\bigg[(l\wedge S_{2}^{-\alpha})^{\mu\nu}v_{1\alpha} - (v_{1}\wedge S_{2}^{-\alpha})^{\mu\nu}l_{\alpha} -\frac{i}{m_{2}}(l\wedge S_{2}^{-\alpha})^{\mu\nu}(l\cdot S^{-}_{1\alpha}) -\frac{i}{m_{2}}(S_{1}^{-\rho}\wedge S_{2}^{-\alpha})^{\mu\nu}\ l_{\rho}l_{\alpha}\bigg]
\end{split}
\end{equation}
Using the eq.\eqref{eq:235} and eq.\eqref{eq:237}, we compute the radiative gauge field for the second $\sqrt{\text{Kerr}_{2}}$ in the soft limit at the subleading order. We obtain
\begin{equation}
\begin{split}
\tilde{J}_{2}^{\mu}(k) & = -\frac{Q_{2}^{2}Q_{1}}{m_{2}}\int \frac{d^{4}l}{(2\pi)^{4}}\ \hat{\delta}(l\cdot v_{1})\hat{\delta}'(l\cdot v_{2}) e^{il\cdot b_{1}} \frac{1}{l^{2}}\\
&\hspace{1cm}\frac{k_{\nu}}{k\cdot v_{2}}(v_{2}\wedge l)^{\mu\nu}\bigg(v_{1}\cdot v_{2} -\frac{i}{m_{2}}(v_{1}\wedge l)_{2} -\frac{i}{m_{1}}(l\wedge v_{2})_{1} +\frac{(l\cdot S_{1}^{-\sigma})(l\cdot S^{-}_{{2}\sigma})}{m_{1}m_{2}}\bigg)\\
&\hspace{1cm}-\frac{Q_{2}^{2}Q_{1}}{m_{2}}\int d\mu(1,2,l) e^{i l\cdot b_{1}}\ \frac{1}{l^{2}}\ \frac{k_{\nu}}{k\cdot v_{2}}\bigg( (v_{2}\wedge v_{1})^{\mu\nu} -\frac{i}{m_{1}}(v_{2}\wedge S_{1}^{-\alpha})^{\mu\nu}l_{\alpha}\bigg)\\
&\hspace{3cm}+i\frac{Q_{2}^{2}Q_{1}}{m_{2}^{2}}\int d\mu(1,2,l)\ \frac{1}{l^{2}}\ e^{il\cdot b_{1}}\\
&\frac{k_{\nu}}{k\cdot v_{2}}\bigg[(l\wedge S_{2}^{-\alpha})^{\mu\nu}v_{1\alpha} - (v_{1}\wedge S_{2}^{-\alpha})^{\mu\nu}l_{\alpha} +\frac{i}{m_{2}}(l\wedge S_{2}^{-\alpha})^{\mu\nu}(l\cdot S^{-}_{1\alpha}) +\frac{i}{m_{2}}(S_{1}^{-\rho}\wedge S_{2}^{-\alpha})^{\mu\nu}\ l_{\rho}l_{\alpha}\bigg]\\
&-i\frac{Q_{2}^{2}Q_{1}}{m_{2}^{2}}\int d\mu(1,2,l)\ e^{il\cdot b_{1}}\ \frac{1}{l^{2}}\ \frac{k_{\nu}}{k\cdot v_{2}}\ \bigg(\frac{k\cdot l}{k\cdot v_{2}}\bigg)\ S_{2}^{\mu\nu}\\
&\hspace{2cm}\bigg(v_{1}\cdot v_{2} -\frac{i}{m_{1}}(v_{2}\wedge l)_{1} -\frac{i}{m_{2}}(l\wedge v_{1})_{2} +\frac{(l\cdot S_{1}^{-\sigma})(l\cdot S^{-}_{{2}\sigma})}{m_{1}m_{2}}\bigg)
\end{split}
\end{equation}
Using eq.\eqref{eq:236} and eq.\eqref{eq:238}, the above expression can be written in a condensed manner as 
\begin{equation}\label{eq:{445}}
\begin{split}
\tilde{J}_{2}^{\mu}(k) & =  \ \frac{Q_{2}}{m_{2}}\frac{k_{\nu}}{k\cdot v_{2}}\Delta S_{2}^{\mu\nu} - \frac{Q_{2}}{m_{2}^{2}}\ \frac{k_{\nu}}{(k\cdot v_{2})^{2}}(k\cdot \Delta p_{2})S_{2}^{\mu\nu}\\
& \hspace{-1cm}-\frac{Q_{2}^{2}Q_{1}}{m_{2}}\frac{k_{\nu}}{k\cdot v_{2}}\bigg(v_{2}\wedge \frac{\p}{\p v_{2}}\bigg)^{\mu\nu}\int d\mu(1,2,l)\ e^{-il\cdot b_{1}}\bigg(v_{1}\cdot v_{2} -\frac{i}{m_{2}}(v_{1}\wedge l)_{2} -\frac{i}{m_{1}}(l\wedge v_{2})_{1} +\frac{(l\cdot S_{1}^{-\sigma})(l\cdot S^{-}_{{2}\sigma})}{m_{1}m_{2}}\bigg)
\end{split}
\end{equation}
The full radiative gauge field till $\mathcal{O}(S_{1}S_{2})$ and to leading order in the coupling is 
 \begin{equation}\label{eq:341}
\begin{split}
\tilde{J}^{\mu}(k) &= \frac{Q_{1}}{m_{1}}\ \bigg(\frac{b_{1}\cdot k}{k\cdot v_{1}}\bigg)\ \frac{k_{\nu}}{k\cdot v_{1}}(v_{1}\wedge \Delta p_{1})^{\mu\nu} +\frac{Q_{1}}{m_{1}}\frac{k_{\nu}}{k\cdot v_{1}}\Delta S_{1}^{\mu\nu} - \frac{Q_{1}}{m_{1}^{2}}\ \frac{k_{\nu}}{(k\cdot v_{1})^{2}}(k\cdot \Delta p_{1})S_{1}^{\mu\nu}\\
& \hspace{-1cm}-\frac{Q_{1}^{2}Q_{2}}{m_{1}}\frac{k_{\nu}}{k\cdot v_{1}}\bigg(v_{1}\wedge \frac{\p}{\p v_{1}}\bigg)^{\mu\nu}\ \int d\mu(1,2,l)\ e^{-il\cdot b_{1}}\bigg(v_{1}\cdot v_{2} -\frac{i}{m_{2}}(v_{1}\wedge l)_{2} -\frac{i}{m_{1}}(l\wedge v_{2})_{1} +\frac{(l\cdot S_{1}^{-\sigma})(l\cdot S^{-}_{{2}\sigma})}{m_{1}m_{2}}\bigg)\\
&+\frac{Q_{2}}{m_{2}}\frac{k_{\nu}}{k\cdot v_{2}}\Delta S_{2}^{\mu\nu} - \frac{Q_{2}}{m_{2}^{2}}\ \frac{k_{\nu}}{(k\cdot v_{2})^{2}}(k\cdot \Delta p_{2})S_{2}^{\mu\nu}\\
& \hspace{-1cm}-\frac{Q_{2}^{2}Q_{1}}{m_{2}}\frac{k_{\nu}}{k\cdot v_{2}}\bigg(v_{2}\wedge \frac{\p}{\p v_{2}}\bigg)^{\mu\nu}\int d\mu(1,2,l)\ e^{-il\cdot b_{1}}\bigg(v_{1}\cdot v_{2} -\frac{i}{m_{2}}(v_{1}\wedge l)_{2} -\frac{i}{m_{1}}(l\wedge v_{2})_{1} +\frac{(l\cdot S_{1}^{-\sigma})(l\cdot S^{-}_{{2}\sigma})}{m_{1}m_{2}}\bigg)
\end{split}
\end{equation}
Before we move to the quantum computation, we match this result with the classical sub leading soft factor. 

\subsection{Classical sub leading soft factor}

In this subsection we shall match the soft limit of the radiative kernel to the classical subleading soft factor for spinning particles till leading order in the low expansion parameter as done in \cite{athira20}. For the first particle we get the expression
\begin{equation}
\begin{split}\label{eq:230}
S^{(1)\mu}(\prod_{i=1}^{2} p_{i},S_{i},k) & = k_{\nu}\frac{J^{\mu\nu}_{1,+}}{p_{1,+}\cdot k} - k_{\nu}\frac{J^{\mu\nu}_{1,-}}{p_{1,-}\cdot k} = k_{\nu}\frac{(L^{\mu\nu}_{1,+} + S^{\mu\nu}_{1,+})}{p_{1,+}\cdot k} - k_{\nu}\frac{(L^{\mu\nu}_{1,-} + S^{\mu\nu}_{1,-})}{p_{1,-}\cdot k}\\
&= \frac{b_{1}\cdot k}{(p_{1}\cdot k)^{2}}(p_{1}\wedge \Delta p_{1})^{\mu\nu} - k_{\nu}S_{1}^{\mu\nu}\ \frac{\Delta p_{1}\cdot k}{(p_{1}\cdot k)^{2}} + \frac{k_{\nu}}{p_{1}\cdot k}\Delta S_{1}^{\mu\nu} + \frac{k_{\nu}}{p_{1}\cdot k}(z_{1}(0)\wedge p_{1})^{\mu\nu}
\end{split}
\end{equation}
where the last term is obtained from the change in the orbital angular momentum with the respect to the origin. In going from the first to the next line we have used
\begin{equation}
\begin{split}
L_{1,+}^{\mu\nu} & = L_{1,-}^{\mu\nu} + (b_{1}\wedge \Delta p_{1})^{\mu\nu} + (z_{1}(0)\wedge p_{1})^{\mu\nu}\\
& = (b_{1}\wedge p_{1})^{\mu\nu} + (b_{1}\wedge \Delta p_{1})^{\mu\nu} + (z_{1}(0)\wedge p_{1})^{\mu\nu}
\end{split}
\end{equation}
We can now compare the two expressions in eq.\eqref{eq:230} and eq.\eqref{eq:438}. We see that apart from the last term there is an exact match between the two expressions. The last term is also consistent as we show in the appendix\ref{B}. Hence, we can write eq.\eqref{eq:230} as
\begin{equation}\label{eq:232}
\begin{split}
&S^{(1)\mu}(\prod_{i=1}^{2}p_{i},S_{i},k) =\frac{b_{1}\cdot k}{(p_{1}\cdot k)^{2}}(p_{1}\wedge \Delta p_{1})^{\mu\nu} - k_{\nu}S_{1}^{\mu\nu}\ \frac{\Delta p_{1}\cdot k}{(p_{1}\cdot k)^{2}} + \frac{k_{\nu}}{p_{1}\cdot k}\Delta S_{1}^{\mu\nu}\\
&-\frac{Q_{1}^{2}Q_{2}}{m_{1}}\frac{k_{\nu}}{k\cdot v_{1}}\bigg(v_{1}\wedge \frac{\p}{\p v_{1}}\bigg)^{\mu\nu}\ \int d\mu(1,2,l)\ e^{-il\cdot b_{1}}\bigg(v_{1}\cdot v_{2} -\frac{i}{m_{2}}(v_{1}\wedge l)_{2} -\frac{i}{m_{1}}(l\wedge v_{2})_{1} +\frac{(l\cdot S_{1}^{-\sigma})(l\cdot S^{-}_{{2}\sigma})}{m_{1}m_{2}}\bigg)\\
\end{split}
\end{equation}
The matching of the subleading soft factor and the soft limit of the radiative gauge field for the second particle proceeds in the same manner as we have outlined for the first particle. Hence, we see that the two expressions match for both the particles. We now turn to the quantum computation of the radiative gauge field for the $\sqrt{\text{Kerr}_{1}}-\sqrt{\text{Kerr}_{2}}$ scattering.

\section{Review of KMOC formalism}\label{sec3}

In this section we shall first review the KMOC formalism for scalar particles and then move to the spinning case. The essential idea is to start with initial wave-packet with asymptotic data peaked around classical values, evolve it with the S-matrix and then compute the expectation value of an observable in this final state. The classical expression is obtained by taking the classical limit. It is taken before any loop integrals are computed which results in computing only a small number of Feynman diagrams, which is the main advantage of the formalism.\\
The initial state is
\begin{equation}
\ket{\Psi} = \int \prod_{i=1}^{2}\ d\Phi(p_{i})\ \phi(p_{i})\ e^{ib_{1}\cdot p_{1}/\hbar}\ \ket{\vec{p}_{1},\vec{p}_{2}},\ \ \ \int d\Phi(p)\ |\phi(p)|^{2} = 1
\end{equation}  
The $\phi(p_{i})$ are gaussian wave functions which have peaks around their classical values, $m_{i}u_{i}$ and the two particles are separated by a distance of $|\vec{b}_{1}|$.
The measure is the usual Lorentz invariant one,
\begin{equation}
d\Phi(p_{i}) = \frac{d^{4}p_{i}}{(2\pi)^{4}}\ \hat{\delta}^{(+)}(p_{i}^{2}-m_{i}^{2}) =  \frac{d^{4}p_{i}}{(2\pi)^{4}}\ \Theta (p_{i}^{0})\ \hat{\delta}(p_{i}^{2}-m_{i}^{2})
\end{equation}
The next step is construct a general observable by a suitable choice of an operator,
\begin{equation}\label{eq:genobs}
\hat{\delta} \hat{O}^{A}(p_{i},\ldots) = \hbar^{\alpha}\big[ \bra{\Psi}\hat{S}\ \hat{O}^{A}\hat{S}\ket{\Psi} - \bra{\Psi}\hat{O}^{A}\ket{\Psi}\big] 
\end{equation}  
The above expression is the change in the expectation value of the operator, $\hat{O}^{A}$, from the initial state. $\hat{S}$ is the S-matrix. The index '$A$' on $\hat{O}^{A}$ could be a colour index or spacetime indices. The dots in the argument in the left hand side denote that the change can also depend on other asymptotic data like colour or spin as we shall see. The $\hbar^{\alpha}$ is a factor to account for the correct dimensions on both sides. The above expression is then recast  by using the unitarity of the $S$-matrix, in terms of two terms, one which depend the $4$-pt amplitude and another which has unitarity cuts stitching tree level amplitudes. The classical limit of the observable is achieved by scaling the couplings, $g\rightarrow g/\sqrt{\hbar}$ and massless exchanged momenta, $q^{\mu} \rightarrow \bar{q}^{\mu}\hbar$. So, we shall be left with an integral over the wavenumber of the field exchanged between the two incoming scattering particles and this is the classical result.\\
The observable of interest to us is the Radiative kernel which in the classical limit gives the radiative gauge / metric field. This observable can be constructed by taking the expectation value of the gauge field operator, $\hat{\mathbb{A}}^{\mu}(x)$\cite{kosower21}. Assuming no incoming radiation in the initial state, the radiative kernel till leading order is 
\begin{equation}\label{eq:336}
\begin{split}
\mathcal{R}^{\mu}(k) & = \braket{\braket{\ \hbar^{3/2}\ \int d\hat{\mu}(1,2,q_{1},q_{2})\ e^{-iq_{1}\cdot b_{1}/\hbar}\ \mathcal{A}_{5}(p_{1},p_{2}\rightarrow \tilde{p}_{1},\tilde{p}_{2},k)\ }}\\
& = \braket{\braket{\ \hbar^{3/2}\ \int d\hat{\mu}(1,2,q_{1},q_{2})\ e^{-iq_{1}\cdot b_{1}/\hbar}\ \hat{\delta}^{(4)}(q_{1}+q_{2}+k)\ A_{5}(p_{1},p_{2}\rightarrow \tilde{p}_{1},\tilde{p}_{2},k)\ }} + \mathcal{O}(T^{\dag}T)
\end{split}
\end{equation}
The $\braket{\braket{\ldots}}$ denotes the coherent state integration, the result of which is plugging in the classical values. The $\tilde{p}_{i} = p_{i}+q_{i}$ and $k$ is four momentum of the outgoing photon. In the classical limit, $k^{\mu} = \hbar \bar{k}^{\mu}$ due to momentum conservation. To leading order in the coupling, only the first term needs to be kept.
As a disclaimer, in practise we shall not use denote put a bar on the wavenumber after taking the classical limit as this makes the expressions more cluttered. However, we will state which steps we take the classical limit to avoid any misunderstanding.\\ 
We review the spinning case now. The formalism was first extended to spinning particles in \cite{MOV}. The observable which gives the classical spin contribution is the Pauli - Lubanski vector. This is motivated by the fact that the Pauli - Lubanski vector is the generator of little group transformations and more importantly, it isolates the spin contribution of any observable. Hence, we have
\begin{equation}\label{eq:337}
s^{\mu}_{ij}(p) = \frac{1}{m}\bra{\vec{p},i}\mathbb{W}^{\mu}\ket{\vec{p},j},\hspace{2cm} s^{\mu} = \sum_{i,j}\int d\Phi(p)\ |\phi(p)|^{2}\ \zeta^{*}_{i}\ s^{\mu}_{ij}(p) \zeta_{j}  
\end{equation}
Here $\zeta_{i}$ can be thought off as a schematic for the coherent state for spin in the little group, $\mathbb{W}^{\mu}$ is the Pauli - Lubanski vector and $\ket{\vec{p},i}$ is Fock space state with little group index `i'. In this note we shall be studying only the lowest order in the classical spin expansion, so we shall not discuss the construction of $\zeta_{i}$. This can be found in \cite{ochirov21}. In this note, we shall think of the classical spin vector as the expectation of the Pauli - Lubanski vector in a coherent state and be agnostic to the details of $\zeta_{i}$.\\
For a generic spinning particle, the $s^{\mu}_{ij}$ can be constructed using the field expansions and the Lorentz representation generator. For $s=1/2$ particle, we get
\begin{equation}
s^{\mu}_{ab} = \frac{\hbar}{4m}\bar{u}_{a}(p)\gamma^{5}\gamma^{\mu}u_{b}(p)
\end{equation}
where $a,b$ are the little group indices on the outgoing and incoming Dirac fermion. The $u^{a}(p)$ is the Dirac spinor and $\gamma^{\mu}$ are the Dirac matrices.\\
For a spinning particle, a general observable (like the linear impulse) will depend on the final and initial $s^{\mu}_{ab}$. Since, the exchange momentum is small the final $s^{\mu}_{ab}$ can be obtained from the initial one by an infinitesimal boost,
\begin{equation}
s^{\mu}_{ab}(p+\hbar\bar{q}) = s^{\mu}_{1ab}(p) + \Delta s_{1ab}^{\mu}(p)
\end{equation}
Finally, the terms linear in $s^{\mu}_{ab}$ will give the contribution of the classical pseudo-vector, $s^{\mu}$ after the use of eq.\eqref{eq:337}. In general for spin-$s$ particle, the classical limit of an observable will contain a maximum of $2s$-pt functions of the Pauli - Lubanski operators\footnote{From an EFT perspective in the matching of the classical limit of the full amplitude, the amplitude contains correlation functions of $\mathbb{W}^{\mu}$ which are identified as contribution of spin operators in the low energy EFT\cite{cfq3.5,bernspin20,porto21}.}. Schematically then in taking the classical result, a general contribution to an observable takes the form
\begin{equation}
(A^{(0)} + A^{(1)}_{\mu_{1}}\braket{s^{\mu_{1}}} + A^{(2)}_{\mu_{1}\mu_{2}}\braket{s^{\mu_{1}}s^{\mu_{2}}} +\ldots + A^{(2s)}_{\mu_{1}\ldots \mu_{2s}}\braket{s^{\mu_{1}}\ldots s^{\mu_{2s}}})
\end{equation}
The first contribution is the scalar term, the second is the linear in spin vector and so on. The expectation value of the spin vectors is taken in the spin coherent states and hence gives the classical result. In \cite{MOV} the classical observables like spin angular impulse and linear impulse for Kerr black holes and the $\sqrt{\text{Kerr}}$ object were obtained for the first few orders in the spin expansion using low finite spin particles.
In the next section we shall follow this same methodology to compute the soft radiative gauge field to the first few orders of spin.

\section{Soft radiative gauge field for $\sqrt{\text{Kerr}_{1}}$-$\sqrt{\text{Kerr}_{2}}$ scattering}\label{sec5}

In this section we shall use the KMOC formalism generalised to include spinning particles to compute classical soft electromagnetic radiation from the scattering of two $\sqrt{\text{Kerr}}$ objects. From the works of \cite{nima19,yutin17}, classical objects like the $\sqrt{\text{Kerr}}$ can be thought of as the classical limit of an infinite spin particle with $S\rightarrow \infty, \hbar\rightarrow 0$ such that $S\hbar=const$, for tree level scattering. This gives the classical result for a $\sqrt{\text{Kerr}}$ to all orders in spin. Since, we need to recover the radiative gauge field for only the first few orders in the spin expansion, we shall work with low finite spin particles. To ascertain the theory to be used, we note that the gyromagnetic ratio for the $\sqrt{\text{Kerr}}$ is $g=2$. The Dirac fermion is the lowest spinning particle with the same gyromagnetic ratio and so, we shall work with it\footnote{The massive spin-1 particle is also another candidate, however, preliminary investigations suggest there is an ambiguity with the classical results obtained with the same.}. It has also been shown that the linear impulse till $\mathcal{O}(S)$ for the $\sqrt{\text{Kerr}}$ can be obtained from the classical limit of the $4$-pt amplitude involving the Dirac fermion\cite{MOV}. For works at $1$-loop on the Kerr black hole and the $\sqrt{\text{Kerr}}$, we refer the reader to \cite{falkowski20,pichini21,aoude22,helset22,menezes22}.\\
We shall consider the scattering of two Dirac fermions. So we start with the radiative kernel to leading order in the coupling, given in eq.\eqref{eq:336}
\begin{equation}\label{eq:449}
\begin{split}
\mathcal{R}^{\mu}(k) & = \braket{\braket{\ \hbar^{3/2}\ \int d\hat{\mu}(1,2,q_{1},q_{2})\ e^{-iq_{1}\cdot b_{1}/\hbar}\ \mathcal{A}_{5}^{\mu\ ab,cd}(p_{1},p_{2}\rightarrow \tilde{p}_{1},\tilde{p}_{2},k)\ }}\\
& = \braket{\braket{\ \hbar^{3/2}\ \int d\hat{\mu}(1,2,q_{1},q_{2})\ e^{-iq_{1}\cdot b_{1}/\hbar}\ \hat{\delta}^{(4)}(q_{1}+q_{2}+k)\ A^{\mu\ ab,cd}_{5}(p_{1},p_{2}\rightarrow \tilde{p}_{1},\tilde{p}_{2},k)\ }}
\end{split}
\end{equation}
The little group indices for the fermion of momentum $\tilde{p}_{1} = p_{1} + q_{1}$ is $a$, for $p_{1}$ it is $b$, for $\tilde{p}_{2} = p_{2} + q_{2}$ its $c$ and for $p_{2}$ it is $d$. We denote the spin angular momentum of the first and second spin-1/2 particle as $S_{1ab}^{\mu\nu}$ and $S_{2cd}^{\mu\nu}$, respectively.\\
The $5$-pt amplitude can be computed using the Feynman rules given in appendix \ref{A} and we get
\begin{equation}
\begin{split}
i & A^{\mu\ ab,cd}_{5}(p_{1},p_{2}\rightarrow \tilde{p}_{1},\tilde{p}_{2},k)\ = \\
& \hspace{-1cm}-iQ^{2}_{1}Q_{2}\bigg[ \bigg(-\frac{p_{1}^{\mu}}{p_{1}\cdot k} + \frac{\tilde{p}^{\mu}_{1}}{\tilde{p}_{1}\cdot k}\bigg)\ \frac{(\bar{u}^{a}(\tilde{p}_{1})\gamma^{\alpha}u^{b}(p_{1}))\ (\bar{v}^{c}(\tilde{p}_{2})\gamma_{\alpha}v^{d}(p_{2}))}{q_{2}^{2}}\\
&\hspace{1cm}+i\ k_{\nu}\bigg(\ \frac{u^{a}(\tilde{p}_{1})\gamma^{\alpha}\hat{\Sigma}^{\mu\nu}u^{b}(p_{1})}{p_{1}\cdot k} - \frac{u^{a}(\tilde{p}_{1})\hat{\Sigma}^{\mu\nu}\gamma^{\alpha}u^{b}(p_{1})}{\tilde{p}_{1}\cdot k}\bigg)\frac{(\bar{v}^{c}(\tilde{p}_{2})\gamma_{\alpha}v^{d}(p_{2}))}{q_{2}^{2}}\bigg]\\
& \hspace{-1cm}-iQ^{2}_{2}Q_{1}\bigg[ \bigg(-\frac{p_{2}^{\mu}}{p_{2}\cdot k} + \frac{\tilde{p}^{\mu}_{2}}{\tilde{p}_{2}\cdot k}\bigg)\ \frac{(\bar{v}^{c}(\tilde{p}_{2})\gamma^{\alpha}v^{d}(p_{2}))\ (\bar{u}^{a}(\tilde{p}_{1})\gamma_{\alpha}u^{b}(p_{1}))}{q_{1}^{2}}\\
&\hspace{1cm}+i\ k_{\nu}\bigg(\ \frac{v^{a}(\tilde{p}_{2})\gamma^{\alpha}\hat{\Sigma}^{\mu\nu}v^{d}(p_{2})}{p_{2}\cdot k} - \frac{v^{c}(\tilde{p}_{2})\hat{\Sigma}^{\mu\nu}\gamma^{\alpha}v^{d}(p_{2})}{\tilde{p}_{2}\cdot k}\bigg)\frac{(\bar{u}^{a}(\tilde{p}_{1})\gamma_{\alpha}u^{b}(p_{1}))}{q_{1}^{2}}\bigg]\\
\end{split}
\end{equation}
where $\hat{\Sigma}^{\mu\nu}$ is the Lorentz group generator in the $s=1/2$ representation. The first two lines are the contribution from the first spin$-1/2$ and the rest is the contribution from the second particle. In writing this we have used the on - shell conditions of the spinors and the Clifford algebra. Plugging this into eq.\eqref{eq:449}, we get
\begin{equation}
\begin{split}
\mathcal{R}^{\mu}(k) & = -\hbar^{3/2}\ Q_{1}^{2}Q_{2}\ \int d\hat{\mu}(1,2,q_{1},q_{2})\ e^{-iq_{1}\cdot b_{1}/\hbar}\ \hat{\delta}^{(4)}(q_{1}+q_{2}+k)\\
&\bigg[ \bigg(-\frac{p_{1}^{\mu}}{p_{1}\cdot k} + \frac{\tilde{p}^{\mu}_{1}}{\tilde{p}_{1}\cdot k}\bigg)\ \frac{(\bar{u}^{a}(\tilde{p}_{1})\gamma^{\alpha}u^{b}(p_{1}))\ (\bar{v}^{c}(\tilde{p}_{2})\gamma_{\alpha}v^{d}(p_{2}))}{q_{2}^{2}}\\
&\hspace{1cm}+i\ k_{\nu}\bigg(\ \frac{u^{a}(\tilde{p}_{1})\gamma^{\alpha}\hat{\Sigma}^{\mu\nu}u^{b}(p_{1})}{p_{1}\cdot k} - \frac{u^{a}(\tilde{p}_{1})\hat{\Sigma}^{\mu\nu}\gamma^{\alpha}u^{b}(p_{1})}{\tilde{p}_{1}\cdot k}\bigg)\frac{(\bar{v}^{c}(\tilde{p}_{2})\gamma_{\alpha}v^{d}(p_{2}))}{q_{2}^{2}}\bigg]\\
&-\hbar^{3/2}\ Q_{2}^{2}Q_{1}\ \int d\hat{\mu}(1,2,q_{1},q_{2})\ e^{-iq_{1}\cdot b_{1}/\hbar}\ \hat{\delta}^{(4)}(q_{1}+q_{2}+k)\\
&\bigg[ \bigg(-\frac{p_{2}^{\mu}}{p_{2}\cdot k} + \frac{\tilde{p}^{\mu}_{2}}{\tilde{p}_{2}\cdot k}\bigg)\ \frac{(\bar{v}^{c}(\tilde{p}_{2})\gamma^{\alpha}v^{d}(p_{2}))\ (\bar{u}^{a}(\tilde{p}_{1})\gamma_{\alpha}u^{b}(p_{1}))}{q_{1}^{2}}\\
&\hspace{1cm}+i\ k_{\nu}\bigg(\ \frac{v^{a}(\tilde{p}_{2})\gamma^{\alpha}\hat{\Sigma}^{\mu\nu}v^{d}(p_{2})}{p_{2}\cdot k} - \frac{v^{c}(\tilde{p}_{2})\hat{\Sigma}^{\mu\nu}\gamma^{\alpha}v^{d}(p_{2})}{\tilde{p}_{2}\cdot k}\bigg)\frac{(\bar{u}^{a}(\tilde{p}_{1})\gamma_{\alpha}u^{b}(p_{1}))}{q_{1}^{2}}\bigg]
\end{split}
\end{equation}
We have suppressed the expectation values in the coherent states here which is $\braket{\braket{\ldots}}$. We shall compute the soft radiative gauge field for the first particle. The contribution is
\begin{equation}
\begin{split}
\mathcal{R}_{1}^{\mu}(k) & = -\hbar^{3/2}\ Q_{1}^{2}Q_{2}\ \int d\hat{\mu}(1,2,q_{1},q_{2})\ e^{-iq_{1}\cdot b_{1}/\hbar}\ \hat{\delta}^{(4)}(q_{1}+q_{2}+k)\\
&\bigg[ \bigg(-\frac{p_{1}^{\mu}}{p_{1}\cdot k} + \frac{\tilde{p}^{\mu}_{1}}{\tilde{p}_{1}\cdot k}\bigg)\ \frac{(\bar{u}^{a}(\tilde{p}_{1})\gamma^{\alpha}u^{b}(p_{1}))\ (\bar{v}^{c}(\tilde{p}_{2})\gamma_{\alpha}v^{d}(p_{2}))}{q_{2}^{2}}\\
&\hspace{1cm}+i\ k_{\nu}\bigg(\ \frac{u^{a}(\tilde{p}_{1})\gamma^{\alpha}\hat{\Sigma}^{\mu\nu}u^{b}(p_{1})}{p_{1}\cdot k} - \frac{u^{a}(\tilde{p}_{1})\hat{\Sigma}^{\mu\nu}\gamma^{\alpha}u^{b}(p_{1})}{\tilde{p}_{1}\cdot k}\bigg)\frac{(\bar{v}^{c}(\tilde{p}_{2})\gamma_{\alpha}v^{d}(p_{2}))}{q_{2}^{2}}\bigg]\\
\end{split}
\end{equation}
We will split this into two types of terms, $\mathcal{R}_{1}^{\mu}(k) = \mathcal{R}_{1a}^{\mu}(k) + \mathcal{R}_{1b}^{\mu}(k)$ where
\begin{equation}\label{eq:453}
\begin{split}
\mathcal{R}_{1a}^{\mu}(k)& = -\hbar^{3/2}\ Q_{1}^{2}Q_{2}\ \int d\hat{\mu}(1,2,q_{1},q_{2})\ e^{-iq_{1}\cdot b_{1}/\hbar}\ \hat{\delta}^{(4)}(q_{1}+q_{2}+k)\\
& \hspace{2cm}\bigg(-\frac{p_{1}^{\mu}}{p_{1}\cdot k} + \frac{\tilde{p}^{\mu}_{1}}{\tilde{p}_{1}\cdot k}\bigg)\ \frac{(\bar{u}^{a}(\tilde{p}_{1})\gamma^{\alpha}u^{b}(p_{1}))\ (\bar{v}^{c}(\tilde{p}_{2})\gamma_{\alpha}v^{d}(p_{2}))}{q_{2}^{2}}\\
\end{split}
\end{equation}
and
\begin{equation}\label{eq:454}
\begin{split}
\mathcal{R}_{1b}^{\mu}(k)& = -\hbar^{3/2}\ Q_{1}^{2}Q_{2}\ \int d\hat{\mu}(1,2,q_{1},q_{2})\ e^{-iq_{1}\cdot b_{1}/\hbar}\ \hat{\delta}^{(4)}(q_{1}+q_{2}+k)\\
&\hspace{2cm}i\ k_{\nu}\bigg(\ \frac{u^{a}(\tilde{p}_{1})\gamma^{\alpha}\hat{\Sigma}^{\mu\nu}u^{b}(p_{1})}{p_{1}\cdot k} - \frac{u^{a}(\tilde{p}_{1})\hat{\Sigma}^{\mu\nu}\gamma^{\alpha}u^{b}(p_{1})}{\tilde{p}_{1}\cdot k}\bigg)\frac{(\bar{v}^{c}(\tilde{p}_{2})\gamma_{\alpha}v^{d}(p_{2}))}{q_{2}^{2}}\\
\end{split}
\end{equation}
We now compute the soft limit of the first term, eq.\eqref{eq:453}. We obtain
\begin{equation}
\begin{split}
\mathcal{R}_{1a}^{\mu}(k)& = -\hbar^{3/2}\ Q_{1}^{2}Q_{2}\ \int d\hat{\mu}(1,2,q_{1},q_{2})\ e^{-iq_{1}\cdot b_{1}/\hbar}\ \sum_{i=1}^{2}k\cdot \frac{\p}{\p q_{i}}\hat{\delta}^{(4)}(q_{1}+q_{2})\\
& \hspace{2cm}\bigg(-\frac{p_{1}^{\mu}}{p_{1}\cdot k} + \frac{\tilde{p}^{\mu}_{1}}{\tilde{p}_{1}\cdot k}\bigg)\ \frac{(\bar{u}^{a}(\tilde{p}_{1})\gamma^{\alpha}u^{b}(p_{1}))\ (\bar{v}^{c}(\tilde{p}_{2})\gamma_{\alpha}v^{d}(p_{2}))}{q_{2}^{2}}\\
&= -\hbar^{3/2}\ Q_{1}^{2}Q_{2}\ \int d\hat{\mu}(1,2,q_{1},q_{2})\ e^{-iq_{1}\cdot b_{1}/\hbar}\ \big[ \hat{\delta}^{(4)}(q_{1}+q_{2}+k) - \hat{\delta}^{(4)}(q_{1}+q_{2})\big]\\
& \hspace{2cm}\bigg(-\frac{p_{1}^{\mu}}{p_{1}\cdot k} + \frac{\tilde{p}^{\mu}_{1}}{\tilde{p}_{1}\cdot k}\bigg)\ \frac{(\bar{u}^{a}(\tilde{p}_{1})\gamma^{\alpha}u^{b}(p_{1}))\ (\bar{v}^{c}(\tilde{p}_{2})\gamma_{\alpha}v^{d}(p_{2}))}{q_{2}^{2}}\\
\end{split}
\end{equation}
We now integrate over with $q_{1} = -q-k$ and $q_{2}=q$ in both the integrals and do a $q\rightarrow -q$ flip to obtain
\begin{equation}
\begin{split}
&\mathcal{R}_{1a}^{\mu}(k) = -Q_{1}^{2}Q_{2}\int \frac{d^{4}q}{(2\pi)^{4}}\ \hat{\delta}(2p_{2}\cdot q - q^{2})\ e^{-iq\cdot b/\hbar}\\
& \bigg\{\ \hat{\delta}(2p_{1}\cdot q +q^{2} -(2p_{1}\cdot k +2q\cdot k))\ e^{ik\cdot b_{1}/\hbar}
 \bigg(-\frac{p_{1}^{\mu}}{p_{1}\cdot k} + \frac{(p_{1}+q)^{\mu}}{(p_{1}+q)\cdot k}\bigg) \frac{\bar{u}^{a}(p_{1}+q-k)\gamma^{\alpha}u^{b}(p_{1})}{q^{2}}\\
&\hspace{1cm} -\hat{\delta}(2p_{1}\cdot q + q^{2})\ \bigg(-\frac{p_{1}^{\mu}}{p_{1}\cdot k} + \frac{(p_{1}+q)^{\mu}}{(p_{1}+q)\cdot k}\bigg) \frac{\bar{u}^{a}(p_{1}+q)\gamma^{\alpha}u^{b}(p_{1})}{q^{2}}\bigg\}(v^{c}(p_{2}-q)\gamma_{\alpha}v^{d}(p_{2}))
\end{split}
\end{equation}
So, keeping only the $\mathcal{O}(1)$ term in the soft limit we obtain
\begin{equation}
\begin{split}
&\mathcal{R}^{\mu}_{1a}(k)= -Q_{1}^{2}Q_{2}\int d\hat{\mu}(1,2,q)\ e^{-iq\cdot b_{1}/\hbar}\\
& \hspace{2cm}i\bigg(\frac{b_{1}\cdot k}{\hbar}\bigg)\ \bigg(-\frac{p_{1}^{\mu}}{p_{1}\cdot k} + \frac{(p_{1}+q)^{\mu}}{(p_{1}+q)\cdot k}\bigg)\ \bar{u}^{a}(p_{1}+q)\gamma^{\alpha}u^{b}(p_{1})\ \frac{(v^{c}(p_{2}-q)\gamma_{\alpha}v^{d}(p_{2}))}{q^{2}}\\
& +Q_{1}^{2}Q_{2}\ \int \frac{d^{4}q}{(2\pi)^{4}}\ \hat{\delta}'(2p_{1}\cdot q +q^{2})\ \hat{\delta}(2p_{2}\cdot q-q^{2})\ e^{-iq\cdot b_{1}/\hbar}\\
& \hspace{2cm}(p_{1}\cdot k + q\cdot k)\ \bigg(-\frac{p_{1}^{\mu}}{p_{1}\cdot k} + \frac{(p_{1}+q)^{\mu}}{(p_{1}+q)\cdot k}\bigg)\ \bar{u}^{a}(p_{1}+q)\gamma^{\alpha}u^{b}(p_{1})\ \frac{(v^{c}(p_{2}-q)\gamma_{\alpha}v^{d}(p_{2}))}{q^{2}}\\
&-Q_{1}^{2}Q_{2}\int d\hat{\mu}(1,2,q)\ e^{-iq\cdot b_{1}/\hbar}\\
 & \hspace{1cm}\bigg(-\frac{p_{1}^{\mu}}{p_{1}\cdot k} + \frac{(p_{1}+q)^{\mu}}{(p_{1}+q)\cdot k}\bigg)\ \bigg(-k\cdot \frac{\p}{\p (p_{1}+q)}\bar{u}^{a}(p_{1}+q)\bigg)\gamma^{\alpha}u^{b}(p_{1})\ \frac{(v^{c}(p_{2}-q)\gamma_{\alpha}v^{d}(p_{2}))}{q^{2}}
\end{split}
\end{equation}
The last term can be evaluated using the Laurent expansion of the spinor
\begin{equation}
\bar{u}_{a}(p_{1}+q) = \bar{u}_{a}(p_{1}) +\frac{p_{1}^{\rho}q_{1}^{\sigma}}{4m_{1}^{2}}\ \bar{u}_{a}(p_{1})[\gamma_{\rho},\gamma_{\sigma}] +\mathcal{O}(q^{2})
\end{equation}  
and then evaluate the derivative as $\frac{\p}{\p \tilde{p}_{1}^{\alpha}}q^{\mu} = \frac{\p}{\p \tilde{p}_{1}^{\alpha}}(\tilde{p}_{1}-p_{1})^{\mu} = \delta^{\mu}_{\alpha}$\footnote{Here $\tilde{p}_{1} = p_{1}+q$.}. Finally, we take the classical limit and we get
\begin{equation}\label{eq:459}
\begin{split}
& \mathcal{R}_{1a}^{\mu}(k) = -iQ_{1}\bigg(\frac{b_{1}\cdot k}{p_{1}\cdot k}\bigg)\int d\mu(1,2,q)\ e^{-iq\cdot b_{1}}\ \frac{k_{\nu}}{p_{1}\cdot k}\ (p_{1}\wedge q)^{\mu\nu}\ A_{4}^{\text{cl}}(p_{1},p_{2}\rightarrow p_{1}+q, p_{2}-q)\\
&+Q_{1}\int \frac{d^{4}q}{(2\pi)^{4}}\ \hat{\delta}'(2p_{1}\cdot q)\hat{\delta}'(2p_{2}\cdot q) e^{-iq\cdot b_{1}}\ \frac{k_{\nu}}{p_{1}\cdot k}\ (p_{1}\wedge q)^{\mu\nu}\ A_{4}^{\text{cl}}(p_{1},p_{2}\rightarrow p_{1}+q, p_{2}-q)\\
& -Q_{1}^{2}Q_{2}\int d\mu(1,2,q)\ e^{-iq\cdot b_{1}}\ \frac{k_{\nu}}{p_{1}\cdot k}(p_{1}\wedge q)^{\mu\nu}\frac{1}{q^{2}}\ \big( -4i(k\wedge p_{2})_{1} + \braket{(k\cdot S^{-\alpha}_{1})(q\cdot S^{-}_{2\alpha})}\big)
\end{split}
\end{equation} 
Here the $\braket{S_{1}S_{2}}$ stand for the $2$-pt function of the two spin tensors for the two particles. The $4$-pt amplitude for the scattering of two Dirac fermions is
\begin{equation}
iA_{4}(p_{1},p_{2}\rightarrow p_{1}+q,p_{2}-q) = -iQ_{1}Q_{2}\frac{(\bar{u}^{a}(\tilde{p}_{1})\gamma^{\alpha}u^{b}(p_{1}))\ (\bar{v}^{c}(\tilde{p}_{2})\gamma_{\alpha}v^{d}(p_{2}))}{q^{2}}
\end{equation}
and the classical limit of which is
\begin{equation}\label{eq:461}
A_{4}^{\text{cl}}(p_{1},p_{2}\rightarrow p_{1}+q,p_{2}-q) = -Q_{1}Q_{2}\frac{4p_{1}\cdot p_{2} + 4i(q\wedge p_{1})_{2} - 4i(q\wedge p_{2})_{1} + 4\braket{(q\cdot S_{1}^{-\alpha})(q\cdot S_{2\alpha}^{-})}}{q^{2}}
\end{equation}
We now evaluate the second term for the first Dirac fermion given in eq.\eqref{eq:454}, which is 
\begin{equation}
\begin{split}
\mathcal{R}_{1b}^{\mu}(k)& = -\hbar^{3/2}\ Q_{1}^{2}Q_{2}\ \int d\hat{\mu}(1,2,q_{1},q_{2})\ e^{-iq_{1}\cdot b_{1}/\hbar}\ \hat{\delta}^{(4)}(q_{1}+q_{2}+k)\\
&\hspace{2cm}i\ k_{\nu}\bigg(\ \frac{u^{a}(\tilde{p}_{1})\gamma^{\alpha}\hat{\Sigma}^{\mu\nu}u^{b}(p_{1})}{p_{1}\cdot k} - \frac{u^{a}(\tilde{p}_{1})\hat{\Sigma}^{\mu\nu}\gamma^{\alpha}u^{b}(p_{1})}{\tilde{p}_{1}\cdot k}\bigg)\frac{(\bar{v}^{c}(\tilde{p}_{2})\gamma_{\alpha}v^{d}(p_{2}))}{q_{2}^{2}}\\
\end{split}
\end{equation}
The soft limit of the above expression is
\begin{equation}
\begin{split}
\mathcal{R}_{1b}^{\mu}(k)& = -\hbar^{3/2}\ Q_{1}^{2}Q_{2}\ \int d\hat{\mu}(1,2,q_{1},q_{2})\ e^{-iq_{1}\cdot b_{1}/\hbar}\ \hat{\delta}^{(4)}(q_{1}+q_{2})\\
&\hspace{2cm}i\ k_{\nu}\bigg(\ \frac{u^{a}(\tilde{p}_{1})\gamma^{\alpha}\hat{\Sigma}^{\mu\nu}u^{b}(p_{1})}{p_{1}\cdot k} - \frac{u^{a}(\tilde{p}_{1})\hat{\Sigma}^{\mu\nu}\gamma^{\alpha}u^{b}(p_{1})}{\tilde{p}_{1}\cdot k}\bigg)\frac{(\bar{v}^{c}(\tilde{p}_{2})\gamma_{\alpha}v^{d}(p_{2}))}{q_{2}^{2}}\\
& = -iQ_{1}^{2}Q_{2}\hbar^{3/2}\ \int d\hat{\mu}(1,2,q_{1},q_{2})\ e^{-iq_{1}\cdot b_{1}/\hbar}\ \hat{\delta}^{(4)}(q_{1}+q_{2})\\&\hspace{5cm}\frac{k_{\sigma}}{p_{1}\cdot k}\ \bar{u}^{a}(\tilde{p}_{1})[\gamma^{\alpha},\hat{\Sigma}^{\mu\sigma}]u^{b}(p_{1}) \frac{(\bar{v}^{c}(\tilde{p}_{2})\gamma_{\alpha}v^{d}(p_{2}))}{q_{2}^{2}}\\
&-iQ_{1}^{2}Q_{2}\hbar^{3/2}\ \int d\hat{\mu}(1,2,q_{1},q_{2})\ e^{-iq_{1}\cdot b_{1}/\hbar}\ \hat{\delta}^{(4)}(q_{1}+q_{2})\\ &\hspace{5cm}\frac{k_{\sigma}}{(p_{1}\cdot k)^{2}}\ \bar{u}^{a}(\tilde{p}_{1})\hat{\Sigma}^{\mu\sigma}\gamma^{\alpha}u^{b}(p_{1})\ k_{\sigma}\ \frac{(\bar{v}^{c}(\tilde{p}_{2})\gamma_{\alpha}v^{d}(p_{2}))}{q_{2}^{2}}
\end{split}
\end{equation}
These two terms can be straightforwardly evaluated using the following identities
\begin{equation}
\begin{split}
&[\gamma^{\alpha},\hat{\Sigma}^{\mu\nu}] = i\ (\eta^{\alpha\mu}\gamma^{\nu} - \eta^{\alpha\nu}\gamma^{\mu})\\
&\hspace{-1.5cm}[\gamma^{\mu},\gamma^{\nu}]\gamma^{\alpha} = 2\eta^{\nu\alpha}\gamma^{\mu} - 2\eta^{\mu\alpha}\gamma^{\nu} - 2i\eps^{\mu\nu\alpha\rho}\gamma_{\rho}\gamma_{5}.
\end{split}
\end{equation}
After taking the classical limit, we get
\begin{equation}
\begin{split}
\mathcal{R}_{1b}^{\mu}(k) &= 4Q_{1}^{2}Q_{2}\ \int d\mu(1,2,q)\ e^{-iq\cdot b_{1}}\ \frac{1}{q^{2}}\ \frac{k_{\nu}}{p_{1}\cdot k}\\
&\big[ (p_{2}\wedge p_{1})^{\mu\nu} + i(p_{2}\wedge S_{1}^{-\sigma})^{\mu\nu}q_{\sigma} - i (S_{2}^{-\rho}\wedge p_{1})^{\mu\nu}q_{\rho} + \braket{(S_{2}^{-\rho}\wedge S_{1}^{-\sigma})^{\mu\nu}}q_{\rho}q_{\sigma}\big]\\
&-iQ_{1}^{2}Q_{2}\ (4m_{1})\ \int d\mu(1,2,q)\ e^{-iq\cdot b_{1}}\ \frac{1}{q^{2}}\ \frac{k_{\nu}}{(p_{1}\cdot k)^{2}}(q\cdot k)\ \eps^{\mu\nu\alpha\rho}(s_{1\rho}\ p_{2\alpha} + i\ \braket{s_{1\rho}S_{2\rho\alpha}}q^{\rho})
\end{split}
\end{equation} 
We now use eq.\eqref{eq:22} to express the spin vector in terms of the spin tensor. We get
\begin{equation}\label{eq:466}
\begin{split}
\mathcal{R}_{1b}^{\mu}(k) &= 4Q_{1}^{2}Q_{2}\ \int d\mu(1,2,q)\ e^{-iq\cdot b_{1}}\ \frac{1}{q^{2}}\ \frac{k_{\nu}}{p_{1}\cdot k}\\
&\big[ -(p_{1}\wedge p_{2})^{\mu\nu} + i(p_{2}\wedge S_{1}^{-\sigma})^{\mu\nu}q_{\sigma} + i (p_{1}\wedge S_{2}^{-\rho})^{\mu\nu}q_{\rho} + \braket{(S_{2}^{-\rho}\wedge S_{1}^{-\sigma})^{\mu\nu}}q_{\rho}q_{\sigma}\big]\\
&-4iQ_{1}^{2}Q_{2}\int d\mu(1,2,q)\ e^{-iq\cdot b_{1}}\ \frac{1}{q^{2}}\ \frac{k_{\nu}}{(p_{1}\cdot k)^{2}}\ (q\cdot k)\\
&\big[(p_{1}\wedge S_{1}^{-\alpha})^{\mu\nu}p_{2\alpha} + (p_{1}\cdot p_{2})\ S_{1}^{\mu\nu} - i\ \braket{(p_{1}\wedge S_{1}^{-\alpha})^{\mu\nu}S_{2\alpha\beta}}q^{\beta} - i\braket{(p_{1}\wedge q)_{2}S_{1}^{\mu\nu}}\big]
\end{split}
\end{equation}
Thus, by adding the eq.\eqref{eq:459} and eq.\eqref{eq:466} we get the soft radiative gauge field for the first particle
\begin{equation}
\begin{split}
\mathcal{R}_{1}^{\mu}(k) & = -Q_{1}\bigg(\frac{b_{1}\cdot k}{p_{1}\cdot k}\bigg)\ \frac{k_{\nu}}{p_{1}\cdot k}\ (p_{1}\wedge \Delta p_{1})^{\mu\nu}\ +Q_{1}\frac{\Delta p_{1}\cdot k}{(p_{1}\cdot k)^{2}}\ k_{\nu}S_{1}^{\mu\nu}\\
& + Q_{1}\frac{k_{\nu}}{p_{1}\cdot k}\bigg(p_{1}\wedge \frac{\p}{\p p_{1}}\bigg)^{\mu\nu}\int\ d\mu(1,2,q)\ e^{-iq\cdot b_{1}}\ \frac{1}{q^{2}}\ A^{\text{cl}}_{4}(1,2,q)\\
&\hspace{1cm}+iQ_{1}^{2}Q_{2}\int d\mu(1,2,q)\ e^{-iq\cdot b_{1}}\ \frac{1}{q^{2}} \frac{k_{\nu}}{p_{1}\cdot k}\\
&\hspace{-1cm}\big[\ (p_{2}\wedge S_{1}^{-\alpha})^{\mu\nu}q_{\alpha}\ - (q\wedge S_{1}^{-\alpha})^{\mu\nu}p_{2\alpha} - i(S_{2}^{-\alpha}\wedge S_{1}^{-\beta})^{\mu\nu}q_{\alpha}q_{\beta}\ - i(q\wedge S_{1}^{-\alpha})^{\mu\nu}(q\cdot S^{-}_{2\ \alpha})\big]  
\end{split}
\end{equation}
where the classical limit of the $4$-pt amplitude is given in eq.\eqref{eq:461} and 
\begin{equation}\label{eq:468}
\Delta p_{1}^{\mu} = i\int d\mu(1,2,q)\ q^{\mu}\ A_{4}^{\text{cl}}(p_{1},p_{2}\rightarrow p_{1}+q,p_{2}-q)
\end{equation}
The soft radiative gauge field for the second particle can be obtained in the exact manner as the first one. We get
\begin{equation}
\begin{split}
\mathcal{R}_{2}^{\mu}(k) & =  Q_{2}\frac{k_{\nu}}{p_{2}\cdot k}\bigg(p_{2}\wedge \frac{\p}{\p p_{2}}\bigg)^{\mu\nu}\int\ d\mu(1,2,q)\ e^{-iq\cdot b_{1}}\ \frac{1}{q^{2}}\ A^{\text{cl}}_{4}(p_{1},p_{2}\rightarrow p_{1}+q,p_{2}-q)\\
&+ Q_{2}\ k_{\nu}\ S_{2}^{\mu\nu}\ \frac{\Delta p_{2}\cdot k}{(p_{2}\cdot k)^{2}}\ +\ iQ_{2}^{2}Q_{1}\int d\mu(1,2,q)\ e^{-iq\cdot b_{1}}\ \frac{1}{q^{2}} \frac{k_{\nu}}{p_{2}\cdot k}\\
&\big[- (p_{1}\wedge S_{2}^{-\alpha})^{\mu\nu}q_{\alpha}\ + (q\wedge S_{2}^{-\alpha})^{\mu\nu}p_{1\alpha} - i(S_{1}^{-\alpha}\wedge S_{2}^{-\beta})^{\mu\nu}q_{\alpha}q_{\beta}\ - i(q\wedge S_{2}^{-\alpha})^{\mu\nu}(q\cdot S^{-}_{1\ \alpha})\big]
\end{split}
\end{equation}
where $\Delta p^{\mu}_{2} = - \Delta p^{\mu}_{1}$. Adding the two radiative gauge field in the soft limit we obtain the full radiative gauge field in the soft limit from KMOC to be
\begin{equation}
\begin{split}
\mathcal{R}^{\mu}(k) & = -Q_{1}\bigg(\frac{b_{1}\cdot k}{p_{1}\cdot k}\bigg)\ \frac{k_{\nu}}{p_{1}\cdot k}\ (p_{1}\wedge \Delta p_{1})^{\mu\nu}\ +Q_{1}\frac{\Delta p_{1}\cdot k}{(p_{1}\cdot k)^{2}}\ k_{\nu}S_{1}^{\mu\nu}\\
& + Q_{1}\frac{k_{\nu}}{p_{1}\cdot k}\bigg(p_{1}\wedge \frac{\p}{\p p_{1}}\bigg)^{\mu\nu}\int\ d\mu(1,2,q)\ e^{-iq\cdot b_{1}}\ \frac{1}{q^{2}}\ A^{\text{cl}}_{4}(1,2,q)\\
&\hspace{1cm}+iQ_{1}^{2}Q_{2}\int d\mu(1,2,q)\ e^{-iq\cdot b_{1}}\ \frac{1}{q^{2}} \frac{k_{\nu}}{p_{1}\cdot k}\\
&\hspace{-1cm}\big[\ (p_{2}\wedge S_{1}^{-\alpha})^{\mu\nu}q_{\alpha}\ - (q\wedge S_{1}^{-\alpha})^{\mu\nu}p_{2\alpha} - i(S_{2}^{-\alpha}\wedge S_{1}^{-\beta})^{\mu\nu}q_{\alpha}q_{\beta}\ - i(q\wedge S_{1}^{-\alpha})^{\mu\nu}(q\cdot S^{-}_{2\ \alpha})\big]\\
&+Q_{2}\frac{k_{\nu}}{p_{2}\cdot k}\bigg(p_{2}\wedge \frac{\p}{\p p_{2}}\bigg)^{\mu\nu}\int\ d\mu(1,2,q)\ e^{-iq\cdot b_{1}}\ \frac{1}{q^{2}}\ A^{\text{cl}}_{4}(1,2,q) + Q_{2}k_{\nu}S_{2}^{\mu\nu}\ \frac{\Delta p_{2}\cdot k}{(p_{2}\cdot k)^{2}}\\
&\hspace{1cm}+iQ_{2}^{2}Q_{1}\int d\mu(1,2,q)\ e^{-iq\cdot b_{1}}\ \frac{1}{q^{2}} \frac{k_{\nu}}{p_{2}\cdot k}\\
&\hspace{-1cm}\big[- (p_{1}\wedge S_{2}^{-\alpha})^{\mu\nu}q_{\alpha}\ + (q\wedge S_{2}^{-\alpha})^{\mu\nu}p_{1\alpha} - i(S_{1}^{-\alpha}\wedge S_{2}^{-\beta})^{\mu\nu}q_{\alpha}q_{\beta}\ - i(q\wedge S_{2}^{-\alpha})^{\mu\nu}(q\cdot S^{-}_{1\ \alpha})\big]
\end{split}
\end{equation}
Using eq.\eqref{eq:468} for $\Delta p_{1}^{\mu}$ and $\Delta p_{2}^{\mu}$, we see that there is an exact match with the classical result in eq.\eqref{eq:341}. Integrating the final result using the integrals given in \cite{KMOC,MOV} we get
\begin{equation}
\begin{split}
&\mathcal{R}^{\mu}(k) = -Q_{1}\bigg(\frac{b_{1}\cdot k}{p_{1}\cdot k}\bigg)\ \frac{k_{\nu}}{p_{1}\cdot k}\ (p_{1}\wedge \Delta p_{1})^{\mu\nu}\ +Q_{1}\frac{\Delta p_{1}\cdot k}{(p_{1}\cdot k)^{2}}\ k_{\nu}S_{1}^{\mu\nu}\\
& -\ Q_{1}^{2}Q_{2}\ \frac{\ln(\omega b_{1})}{2\pi}\frac{m_{1}^{2}m_{2}^{2}}{\{(p_{1}\cdot p_{2})^{2}-m_{1}^{2}m_{2}^{2}\}^{3/2}}\ \frac{k_{\nu}}{p_{1}\cdot k}(p_{1}\wedge p_{2})^{\mu\nu}\\
& +\frac{Q_{1}^{2}Q_{2}}{2\pi}\ \frac{1}{b_{1}^{2}}\ \frac{(p_{1}\cdot p_{2})}{\{(p_{1}\cdot p_{2})^{2}-m_{1}^{2}m_{2}^{2}\}^{3/2}}\ \frac{k_{\nu}}{p_{1}\cdot k}(p_{1}\wedge p_{2})^{\mu\nu}\\
&\hspace{1cm}\big[ (b_{1}\wedge p_{2})_{1} + (p_{1}\wedge b_{1})_{2} - (S_{1}\cdot S_{2}) + \frac{2}{b_{1}^{2}}\ (b_{1}\cdot S_{1}^{-\alpha})(b_{1}\cdot S^{-}_{2\alpha})\big]\\
&-\ \frac{Q_{1}^{2}Q_{2}}{2\pi}\frac{1}{b_{1}^{2}}\frac{1}{\sqrt{(p_{1}\cdot p_{2})^{2}-m_{1}^{2}m_{2}^{2}}}\ \frac{k_{\nu}}{p_{1}\cdot k}\ (p_{1}\wedge S_{2}^{-\alpha})^{\mu\nu}b_{1\alpha}\\
& -\frac{Q_{1}^{2}Q_{2}}{2\pi}\ \frac{1}{b_{1}^{2}}\frac{1}{\sqrt{(p_{1}\cdot p_{2})^{2}-m_{1}^{2}m_{2}^{2}}}\ \frac{k_{\nu}}{p_{1}\cdot k}\ \big[\ (p_{2}\wedge S_{1}^{\alpha})^{\mu\nu}b_{1\alpha} - (b_{1}\wedge S_{1}^{-\alpha})^{\mu\nu}p_{2\alpha}\big]\\
&+\frac{Q_{1}^{2}Q_{2}}{\pi}\ \frac{1}{b_{1}^{4}}\frac{1}{\sqrt{(p_{1}\cdot p_{2})^{2}-m_{1}^{2}m_{2}^{2}}}\ \frac{k_{\nu}}{p_{1}\cdot k}\ \big[(S_{2}^{-\alpha}\wedge S_{1}^{-\beta})^{\mu\nu}b_{1\alpha}b_{1\beta} + (b_{1}\wedge S_{1}^{-\alpha})^{\mu\nu}(b_{1}\cdot S^{-}_{2\alpha})\big]\\
& -\frac{Q_{1}^{2}Q_{2}}{\pi}\ \frac{1}{b_{1}^{2}}\frac{1}{\sqrt{(p_{1}\cdot p_{2})^{2}-m_{1}^{2}m_{2}^{2}}}\ \frac{k_{\nu}}{p_{1}\cdot k}\ (S_{2}^{-\alpha}\wedge S^{-}_{1\alpha})^{\mu\nu}\\
& -\frac{Q_{1}^{2}Q_{2}}{2\pi}\ \frac{1}{b_{1}^{2}}\frac{1}{\{(p_{1}\cdot p_{2})^{2}-m_{1}^{2}m_{2}^{2}\}^{3/2}}\ \frac{k_{\nu}}{p_{1}\cdot k}\\
& \big[ - (S^{-\alpha}_{2}\wedge S^{-\beta}_{1})^{\mu\nu}\ p_{1\alpha}p_{2\beta}\ (p_{1}\cdot p_{2}) - (p_{1}\cdot p_{2})\ (p_{2}\wedge S_{1}^{-\alpha})^{\mu\nu}(p_{1}\cdot S^{-}_{2\alpha}) +\ m_{2}^{2}(p_{1}\wedge S_{1}^{-\alpha})^{\mu\nu}(p_{1}\cdot S^{-}_{2\alpha}) \big]\\
& +Q_{2}\frac{\Delta p_{2}\cdot k}{(p_{2}\cdot k)^{2}}\ k_{\nu}S_{2}^{\mu\nu} -\ Q_{2}^{2}Q_{1}\ \frac{\ln(\omega b_{1})}{2\pi}\frac{m_{1}^{2}m_{2}^{2}}{\{(p_{1}\cdot p_{2})^{2}-m_{1}^{2}m_{2}^{2}\}^{3/2}}\ \frac{k_{\nu}}{p_{2}\cdot k}(p_{2}\wedge p_{1})^{\mu\nu}\\
& +\frac{Q_{2}^{2}Q_{1}}{2\pi}\ \frac{1}{b_{1}^{2}}\ \frac{(p_{1}\cdot p_{2})}{\{(p_{1}\cdot p_{2})^{2}-m_{1}^{2}m_{2}^{2}\}^{3/2}}\ \frac{k_{\nu}}{p_{2}\cdot k}(p_{2}\wedge p_{1})^{\mu\nu}\\
&\hspace{1cm}\big[ (b_{1}\wedge p_{2})_{1} + (p_{1}\wedge b_{1})_{2} - (S_{1}\cdot S_{2}) + \frac{2}{b_{1}^{2}}\ (b_{1}\cdot S_{1}^{-\alpha})(b_{1}\cdot S^{-}_{2\alpha})\big]\\
&+\ \frac{Q_{2}^{2}Q_{1}}{2\pi}\frac{1}{b_{1}^{2}}\frac{1}{\sqrt{(p_{1}\cdot p_{2})^{2}-m_{1}^{2}m_{2}^{2}}}\ \frac{k_{\nu}}{p_{2}\cdot k}\ (p_{2}\wedge S_{1}^{-\alpha})^{\mu\nu}b_{1\alpha}\\
& -\frac{Q_{2}^{2}Q_{1}}{2\pi}\ \frac{1}{b_{1}^{2}}\frac{1}{\sqrt{(p_{1}\cdot p_{2})^{2}-m_{1}^{2}m_{2}^{2}}}\ \frac{k_{\nu}}{p_{2}\cdot k}\ \big[\ -(p_{1}\wedge S_{2}^{\alpha})^{\mu\nu}b_{1\alpha} - (b_{1}\wedge S_{2}^{-\alpha})^{\mu\nu}p_{1\alpha}\big]\\
&+\frac{Q_{2}^{2}Q_{1}}{\pi}\ \frac{1}{b_{1}^{4}}\frac{1}{\sqrt{(p_{1}\cdot p_{2})^{2}-m_{1}^{2}m_{2}^{2}}}\ \frac{k_{\nu}}{p_{2}\cdot k}\ \big[(S_{1}^{-\alpha}\wedge S_{2}^{-\beta})^{\mu\nu}b_{1\alpha}b_{1\beta} + (b_{1}\wedge S_{2}^{-\alpha})^{\mu\nu}(b_{1}\cdot S^{-}_{1\alpha})\big]\\
& -\frac{Q_{2}^{2}Q_{1}}{\pi}\ \frac{1}{b_{1}^{2}}\frac{1}{\sqrt{(p_{1}\cdot p_{2})^{2}-m_{1}^{2}m_{2}^{2}}}\ \frac{k_{\nu}}{p_{2}\cdot k}\ (S_{1}^{-\alpha}\wedge S^{-}_{2\alpha})^{\mu\nu}\\
& -\frac{Q_{2}^{2}Q_{1}}{2\pi}\ \frac{1}{b_{1}^{2}}\frac{1}{\{(p_{1}\cdot p_{2})^{2}-m_{1}^{2}m_{2}^{2}\}^{3/2}}\ \frac{k_{\nu}}{p_{2}\cdot k}\\
& \big[ - (S^{-\alpha}_{1}\wedge S^{-\beta}_{2})^{\mu\nu}\ p_{1\beta}p_{2\alpha}\ (p_{1}\cdot p_{2}) - (p_{1}\cdot p_{2})\ (p_{1}\wedge S_{2}^{-\alpha})^{\mu\nu}(p_{2}\cdot S^{-}_{1\alpha}) +\ m_{1}^{2}(p_{2}\wedge S_{2}^{-\alpha})^{\mu\nu}(p_{2}\cdot S^{-}_{1\alpha}) \big]
\end{split}
\end{equation}
This is the final result. We conclude this section with a few remarks.
\begin{itemize}
\item We can identify the terms in the above terms with the various contributions in the classical sub leading soft factor, in eq.\eqref{eq:230}. The two terms in the first line are the contributions to the radiative gauge field coming from a part of the change in the orbital angular momentum and the linear impulse. The terms from the second line to the eight line arise from the $(z_{1}(0)\wedge p_{1})^{\mu\nu}$ term in the orbital angular momentum. The terms in the ninth and tenth line arise from the change in the spin tensor of the first particle. Then the next line has a contribution from the linear impulse of the second particle and till the second last line, they again arise from the corresponding $(z_{2}(0)\wedge p_{2})^{\mu\nu}$ for the second particle. The second last and last line come from the change in the spin angular momentum of the second particle.
\item Just like scalar case\cite{athira20}, the above result is consistent with the classical sub-leading soft factor in $D=4$\cite{sahoo19} and we also see that the $\ln\omega$ term is independent of the spin of the particle. As mentioned in \cite{athira20}, this is because the $\ln\omega$ term comes from the integration region $\omega \ll |\vec{l}|^{-1} \ll |\vec{a}|\ll |\vec{b}_{1}|$. 
\end{itemize} 

\section{Conclusions}

In this short note, we took the first steps to proving the classical subleading soft photon theorem for spinning particles, for the $\sqrt{\text{Kerr}}$ particle, from the quantum subleading soft photon theorem. We did this by using the KMOC formalism generalised to describe classical spinning particles. We took the soft limit of the radiative kernel for $\sqrt{\text{Kerr}_{1}}-\sqrt{\text{Kerr}_{2}}$ scattering and showed that in the classical limit it reproduces the classical subleading soft photon theorem till $\mathcal{O}(S_{1}S_{2})$ and to LO in the coupling.\\
The obvious issue would be to extend this perturbative proof of the subleading soft photon theorem at leading order in the coupling, to all orders in spin and also study the analogous case of low energy gravitational radiation for the Kerr black hole. It would also be interesting to extend this perturbative proof to higher orders in the coupling. For spinning particles in gravity, the leading spin dependent non-analytic term in $D=4$ has also been derived in \cite{sahoo21}. It would be interesting to relate this to the subsubleading soft factor in gravity.

\section{Acknowledgments}

We would like to thank Alok Laddha for the many fruitful discussions during the project and for his suggestions on the manuscript. We would also like to thank Arkajyoti Manna and Samim Akhtar for discussions on some conceptual points during the course of the project.

\appendix

\section{Matching of the $S^{(1)\mu}$ with the soft limit of the radiative gauge field}\label{B}

So, the starting expression is the acceleration of the first particle given in eq.\eqref{eq:228}. We integrate twice and put $\tau=0$ to get
\begin{equation}
\begin{split}
&z_{1}^{\mu}(0) = -i\frac{Q_{1}Q_{2}}{m_{1}}\int \frac{d^{4}l}{(2\pi)^{4}}\ \hat{\delta}(l\cdot v_{2})\ e^{-il\cdot b_{1}}\ \frac{1}{l^{2}}\frac{1}{(l\cdot v_{1})^{2}_{+}}\\
&\bigg[ l^{\mu}(v_{1}\cdot v_{2})-\ (l\cdot v_{1})v_{2}^{\mu} -\frac{i}{m_{2}}\ \{l^{\mu}(v_{1}\wedge l)_{2} -\ (l\cdot v_{1})\ S^{\mu\alpha}_{2} l_{\alpha}\} + l^{\mu}\bigg(-\frac{i}{m_{1}}(l\wedge v_{2})_{1} +\ \frac{(l\cdot S_{1}^{-\alpha})(l\cdot S_{2\alpha}^{-})}{m_{1}m_{2}}\bigg)\bigg]
\end{split}
\end{equation}
Here $(l\cdot v_{1})_{+} = l\cdot v_{1}+i\epsilon$. The $i\epsilon$ prescription comes from requiring the initial condition that initially the particle is a free particle that is, $z_{1}^{\mu}(\tau\rightarrow -\infty)\rightarrow 0$ as discussed in \cite{gb1}. So,
\begin{equation}
\begin{split}
&(z_{1}(0)\wedge p_{1})^{\mu\nu} = iQ_{1}Q_{2}\int \frac{d^{4}l}{(2\pi)^{4}}\ \hat{\delta}(l\cdot v_{2})\ e^{-il\cdot b_{1}}\ \frac{1}{l^{2}}\frac{1}{(l\cdot v_{1})^{2}_{+}}\\
&\bigg[(v_{1}\wedge v_{2})^{\mu\nu}(l\cdot v_{1}) - (v_{1}\wedge l)^{\mu\nu}(v_{1}\cdot v_{2}) + \frac{i}{m_{2}}\{ (v_{1}\wedge l)^{\mu\nu}\ (v_{1}\wedge l)_{2} - (l\cdot v_{1})\ (v_{1}\wedge S^{-\alpha}_{2})^{\mu\nu}l_{\alpha}\}\\
&\hspace{6.5cm}-(v_{1}\wedge l)^{\mu\nu}\bigg(-\frac{i}{m_{1}}(l\wedge v_{2})_{1} +\ \frac{(l\cdot S_{1}^{-\alpha})(l\cdot S_{2\alpha}^{-})}{m_{1}m_{2}}\bigg)\bigg]\\
&=iQ_{1}Q_{2}\bigg(v_{1}\wedge \frac{\p}{\p v_{1}}\bigg)^{\mu\nu}\ \int \frac{d^{4}l}{(2\pi)^{4}}\ \hat{\delta}(l\cdot v_{2})\ e^{-il\cdot b_{1}}\ \frac{1}{l^{2}}\frac{1}{(l\cdot v_{1})_{+}}\\
&\hspace{6.5cm}\bigg(v_{1}\cdot v_{2} -\frac{i}{m_{2}}(v_{1}\wedge l)_{2} -\frac{i}{m_{1}}(l\wedge v_{2})_{1} +\frac{(l\cdot S_{1}^{-\sigma})(l\cdot S^{-}_{{2}\sigma})}{m_{1}m_{2}}\bigg)
\end{split}
\end{equation}
We shall now do the partially do the integrals involved in the above expression to show that this matches with the expression in eq.\eqref{eq:232}. There are four types of integrals we need to study. The first of which is 
\begin{equation}
I_{1} = \int \frac{d^{4}l}{(2\pi)^{4}}\ \hat{\delta}(l\cdot v_{2})\ e^{-il\cdot b_{1}}\ \frac{1}{(l\cdot v_{1})_{+}}\ \frac{1}{l^{2}}
\end{equation}
We choose the frame in which the first particle travels in the $z$- direction and the second particle is at rest,
\begin{equation}\label{eq:b83}
v_{2}^{\mu} = (1,\vec{0})\ \ \ \ \ \, v_{1}^{\mu} =\gamma(1,0,0,\beta).
\end{equation}
In this frame we can do the $l^{0}$ integral and we get
\begin{equation}
I_{1} = \frac{1}{\gamma\beta}\int \frac{d^{2}l}{(2\pi)^{2}}\ e^{i\vec{l}_{\perp}\cdot \vec{b}_{1}}\int \frac{dl}{(2\pi)}\ \frac{1}{\vec{l}^{2}_{\perp}+\ l^{2}}\frac{1}{l-i\epsilon}
\end{equation}
where $l^{3} = l$. Since the impact parameter is perpendicular to both initial velocities, it comes out of the integral. The above integral can be done using a contour which closes in the lower half plane and we get
\begin{equation}
I_{1} = -\frac{i}{\gamma\beta}\int \frac{d^{2}l_{\perp}}{(2\pi)^{2}}\ e^{i\vec{l}_{\perp}\cdot \vec{b}_{1}}\ \frac{1}{l_{\perp}^{2}}
\end{equation}
Now it can be shown quite easily by going to the frame in eq.\eqref{eq:b83} that 
\begin{equation}
I_{1} = i\int \frac{d^{4}l}{(2\pi)^{4}}\ \hat{\delta}(l\cdot v_{1})\hat{\delta}(l\cdot v_{2})\ e^{-il\cdot b_{1}}\ \frac{1}{l^{2}}
\end{equation}
The second integral we have is
\begin{equation}
I_{2} = -\frac{i}{m_{2}}\int \frac{d^{4}l}{(2\pi)^{4}}\ \hat{\delta}(l\cdot v_{2})\ e^{-il\cdot b_{1}}\ \frac{1}{{l}^{2}}\frac{(v_{1}\wedge l)_{2}}{(l\cdot v_{1})_{+}}
\end{equation}
In the frame we have chosen, we have $S_{2}^{\mu 0} = 0$ coming from the SSC of the second particle. Since, the first particle travels in the $z$-direction, this means that $(v_{1}\wedge l)_{2}$ only depends on the $(x,y)$ coordinates and thus by doing the $l^{0}$ and $l^{3}$ integrals as done in the previous integral we get
\begin{equation}
I_{2} = \frac{1}{m_{2}} \int \frac{d^{4}l}{(2\pi)^{4}}\ \hat{\delta}(l\cdot v_{1})\hat{\delta}(l\cdot v_{2})\ e^{-il\cdot b_{1}}\ \frac{(v_{1}\wedge l)_{2}}{l^{2}}
\end{equation}
Similarly the further two integrals also can be evaluated by arguing that only the $l$-integrals which depend on $l_{perp}$ give non-zero answers and we get
\begin{equation}
I_{3} = -\frac{i}{m_{1}}\int \frac{d^{4}l}{(2\pi)^{4}}\ \hat{\delta}(l\cdot v_{2})\ e^{-il\cdot b_{1}}\ \frac{1}{{l}^{2}}\frac{(l\wedge v_{2})_{1}}{(l\cdot v_{1})_{+}} = \frac{1}{m_{1}} \int \frac{d^{4}l}{(2\pi)^{4}}\ \hat{\delta}(l\cdot v_{1})\hat{\delta}(l\cdot v_{2})\ e^{-il\cdot b_{1}}\ \frac{(l\wedge v_{2})_{1}}{l^{2}}
\end{equation}
and finally
\begin{multline}
I_{4} = \int \frac{d^{4}l}{(2\pi)^{4}}\ \hat{\delta}(l\cdot v_{2})\ e^{-il\cdot b_{1}}\ \frac{1}{{l}^{2}}\frac{(l\cdot S_{1}^{-\alpha})(l\cdot S_{2\alpha}^{-})}{m_{1}m_{2}}\frac{1}{(l\cdot v_{1})_{+}}\\
= i \int \frac{d^{4}l}{(2\pi)^{4}}\ \hat{\delta}(l\cdot v_{1})\hat{\delta}(l\cdot v_{2})\ e^{-il\cdot b_{1}}\ \frac{1}{{l}^{2}}\frac{(l\cdot S_{1}^{-\alpha})(l\cdot S_{2\alpha}^{-})}{m_{1}m_{2}}
\end{multline}
So, taking account all these integrals it can readily be checked that
\begin{equation}
\begin{split}
&(z_{1}(0)\wedge p_{1})^{\mu\nu} =-Q_{1}Q_{2}\bigg(v_{1}\wedge \frac{\p}{\p v_{1}}\bigg)^{\mu\nu}\ \int \frac{d^{4}l}{(2\pi)^{4}}\ \hat{\delta}(l\cdot v_{1})\hat{\delta}(l\cdot v_{2})\ e^{-il\cdot b_{1}}\ \frac{1}{l^{2}}\\
&\hspace{6.5cm}\bigg(v_{1}\cdot v_{2} -\frac{i}{m_{2}}(v_{1}\wedge l)_{2} -\frac{i}{m_{1}}(l\wedge v_{2})_{1} +\frac{(l\cdot S_{1}^{-\sigma})(l\cdot S^{-}_{{2}\sigma})}{m_{1}m_{2}}\bigg)
\end{split}
\end{equation}
Plugging this expression into eq.\eqref{eq:230} we get eq.\eqref{eq:232}.

\section{Notations and Feynman rules}\label{A}

We shall define some of the notations in this appendix that we use in the paper. Most of the expressions will involve the spin tensor and to simplify our lives we use the following notations:
\begin{equation}\label{eq:11}
\begin{split}
&(a\wedge b)^{\mu\nu} = a^{\mu}b^{\nu}-a^{\nu}b^{\mu},\\
&(a\wedge b)_{1} = a_{\alpha}S^{\alpha\beta}b_{\beta},\\
&a\cdot S^{-\rho} = a_{\alpha}S^{\alpha\rho},\\
&(a\wedge S^{-\rho})^{\mu\nu} (b\cdot S^{-}_{\ \rho}) = (a^{\mu}S^{\nu\rho} - a^{\nu}S^{\mu\rho})\ b_{\alpha}S^{\alpha}_{\ \rho}
\end{split}
\end{equation}
The various measures are denoted by 
\begin{equation}
\begin{split}
&d\mu(1,2,q_{1},q_{2}) = \prod_{i=1}^{2}\ d^{4}q_{i}\ \hat{\delta}(2p_{i}\cdot q_{i})\\
&d\mu(1,2,q) = d^{4}q\ \hat{\delta}(2p_{1}\cdot q)\ \hat{\delta}(2p_{2}\cdot q)\\
&d\hat{\mu}(1,2,q_{1},q_{2}) = \prod_{i=1}^{2}\ d^{4}q_{i}\ \hat{\delta}(2p_{i}\cdot q_{i} + q_{i}^{2})
\end{split}
\end{equation}
 The Feynman rules for the Dirac fermion are the following
\begin{align}
\text{interaction vertex}\ :\ -ie\gamma^{\mu}\\
\text{propagator}\ :\ i\frac{(\slashed{p}+m)}{p^{2}-m^{2}}
\end{align}
and the Clifford algebra is
\begin{equation}
\{\gamma^{\mu},\gamma^{\nu}\} = 2\eta^{\mu\nu}\ \ ,\ \ \ \ \ \ \\
\hat{\Sigma}^{\mu\nu} = \frac{i}{4}[\gamma^{\mu},\gamma^{\nu}].
\end{equation}
The photon propagator is
\begin{equation}
\text{photon propagator}\ :\ i\frac{\eta^{\mu\nu}}{q^{2}}
\end{equation}

\end{document}